\documentclass[conference]{IEEEtran}
\IEEEoverridecommandlockouts

\usepackage{amsmath,amssymb,amsfonts}
\usepackage{algpseudocode}
\usepackage{algorithm}

\usepackage{cite}

\usepackage{graphicx}
\usepackage{textcomp}
\usepackage{xcolor}
\def\BibTeX{{\rm B\kern-.05em{\sc i\kern-.025em b}\kern-.08em
    T\kern-.1667em\lower.7ex\hbox{E}\kern-.125emX}}
\usepackage{booktabs}
 
\usepackage{comment}
\usepackage{multirow, graphicx}
\usepackage{subcaption}
\usepackage[a-1b]{pdfx}
\usepackage{hyperref}
\usepackage{amsthm}
\usepackage{comment}
\usepackage{enumitem}
\setlist[itemize]{leftmargin=*}
\setlist[enumerate]{leftmargin=*}

\usepackage{microtype}
\SetExpansion
 [ context = sloppy,
 stretch = 30,
 shrink = 60,
 step = 5 ]
 { encoding = {OT1,T1,TS1} }
 { }
\microtypecontext{expansion=sloppy}

\usepackage{balance}

\usepackage{courier}
\usepackage[font=bf,skip=0pt]{caption}
\usepackage{subcaption}
\usepackage{soul,color}
\usepackage{textcomp}
\usepackage{xcolor}

\usepackage{xspace}

\newcommand{\model}{LearnedWMP\xspace}

\newcommand{\stitle}[1]{\vspace{1ex}\noindent{\bf #1}}

\begin{document}

\title{LearnedWMP: Workload Memory Prediction Using Distribution of Query Templates}

\makeatletter

\newcommand{\linebreakand}{%
  \end{@IEEEauthorhalign}
  \hfill\mbox{}\par
  \mbox{}\hfill\begin{@IEEEauthorhalign}
}
\makeatother

\IEEEaftertitletext{\vspace{-1\baselineskip}} 

\author{\IEEEauthorblockN{Shaikh Quader}
\IEEEauthorblockA{York University, Canada}
\and
\IEEEauthorblockN{Andres Jaramillo}
\IEEEauthorblockA{York University, Canada}

\and
\IEEEauthorblockN{Sumona Mukhopadhyay}
\IEEEauthorblockA{California Polytechnic State Univ., USA} 
\and
\IEEEauthorblockN{Ghadeer Abuoda}
\IEEEauthorblockA{York University, Canada }

\linebreakand 

\IEEEauthorblockN{Calisto Zuzarte}
\IEEEauthorblockA{IBM Canada Ltd. }
\and
\IEEEauthorblockN{David Kalmuk}
\IEEEauthorblockA{IBM Canada Ltd. }
\and
\IEEEauthorblockN{Marin Litoiu}
\IEEEauthorblockA{York University, Canada}
\and
\IEEEauthorblockN{Manos Papagelis}
\IEEEauthorblockA{York University, Canada}
}

\maketitle

\begin{abstract}
In a modern DBMS, {\em working memory} is frequently the limiting factor when processing in-memory analytic query operations such as joins, sorting, and aggregation. Existing resource estimation approaches for a DBMS estimate the resource consumption of a query by computing an estimate of each individual database operator in the query execution plan. Such an approach is slow and error-prone as it relies upon simplifying assumptions, such as uniformity and independence of the underlying data. Additionally, the existing approach focuses on individual queries separately and does not factor in other queries in the workload that may be executed concurrently. 
In this research, we are interested in query performance optimization under concurrent execution of {\em a batch of queries} ({\em a workload}). Specifically, we focus on predicting the memory demand for a workload rather than providing separate estimates for each query within it. 
We introduce the problem of {\em workload memory prediction} and formalize it as a {\em distribution regression} problem. We propose Learned Workload Memory Prediction (LearnedWMP) to improve and simplify estimating the working memory demands of workloads. The LearnedWMP model groups queries based on their similarity in query plans. Each query group is referred to as a {\em query template}. For an input workload, LearnedWMP generates a histogram representation of the query templates of all queries in the workload and uses a distribution regressor to predict the workload's memory demand. Through a comprehensive experimental evaluation, we show that LearnedWMP reduces the memory estimation error of the state-of-the-practice method by up to 47.6\%. Compared to an alternative single-query model, during training and inferencing, the LearnedWMP model and its variants were 3x to 10x faster. Moreover, LearnedWMP-based models were at least 50\% smaller in most cases. Overall, the results demonstrate the advantages of the LearnedWMP approach and its potential for a broader impact on query performance optimization.
\end{abstract}


\section{Introduction}
\label{sec:intro}
Estimating resource usage of database queries is critical to many database operations and decision-making tasks, such as admission control, workload management, and capacity planning~\cite{ganapathi2009predicting}. {\em Working memory} is a region of the system memory where DBMS performs in-memory operations, such as sort and aggregation while executing queries. 
Using a limited system memory, the DBMS can only execute a finite number of in-memory operations (i.e., queries) simultaneously. Inaccurate estimation of a query's working memory requirement causes the DBMS to either under- or over-commit the memory. This hinders the DBMS from achieving optimal query performance, which includes faster query execution and higher throughput, and could potentially result in query failures as well. To achieve high query performance, the DBMS needs accurate query working memory estimations before admitting them for execution.  

\stitle{The State of the Practice \& Limitations.}
 In modern DBMS, the query optimizer's cost model tends to derive estimations for each query independently, missing the opportunity to leverage insights from batches of queries that collectively stress specific resource utilization, such as memory \cite{ganapathi2009predicting}. For example, concurrently processing two queries with each including a \texttt{group by} operation can require higher working memory than executing each query separately. If this collective memory requirement surpasses available system memory, it can lead to increased query execution times, decreasing the DBMS's throughput and query failure. 
As another option, creating a general cost model that is not tied to specific resources and workloads poses a challenge because of the diverse varieties in the database schema, query structures, and data distributions. Despite dedicated efforts over the past decade, the complexity arising from these variations makes the task difficult (e.g.,\cite{li2012robust,akdere2012learning,paul2021database,siddiqui2020cost,sun2019end,wu2013towards,zhao2022queryformer,wu2014uncertainty}). In current practice, human experts traditionally define features and rules for calculating runtime metrics of queries, but this is an expensive and non-generalizable approach \cite{marcus2019plan}. 
To our knowledge, there is currently no practical engine or optimizer capable of predicting with high accuracy memory requirement for a batch of queries. Predicting memory needs for a \emph{batch of queries} simultaneously carries the premise of delivering a more precise estimate of the overall memory demand, helping to avoid excessive or insufficient memory resource allocation, thereby enhancing system stability and performance. This necessitates the development of specialized memory prediction techniques designed for query batches, enabling the generation of precise memory estimates suitable for integration into the DBMS \cite{han2021cardinality}.

\stitle{Our Approach}. We propose a novel approach for estimating the memory demand of a batch of database queries, a {\em workload}. Our approach is a departure from the existing approach of estimating the resource usage of each query separately \cite{han2021cardinality, hilprecht2019deepdb}, especially for cardinality estimation \cite{kim2022learned,yang2020neurocard, liu2015cardinality, hilprecht2019deepdb, kipf2018learned, hasan2020deep}, which is a distinct task from working memory estimation, but often lead to inaccuracies, contributing to imprecise memory estimations \cite{leis2017cardinality,leis2015good}. 
We exploit the observation that DBMS executes queries in batches and that workload queries often have competitive resource demands. By modeling the resource demand of a batch of queries, we expect to achieve higher accuracy in estimating resources. Also, we expect our approach will reduce the development and maintenance costs of the DBMS's resource estimator and speed up the computation of resource estimations. As an embodiment of our idea, we initially focused on estimating the working memory of database workloads. We design a \underline{Learned} \underline{W}orkload \underline{M}emory \underline{P}rediction (LearnedWMP) model in three steps. {\em First}, we use an intuition that queries with similar plan characteristics and estimated cardinalities have similar memory demand. Based on this intuition, we use historical queries to learn {\em query templates} that serve as groups for queries with similar memory demands. {\em Second}, we randomly divide training queries into fixed-size training workloads and represent each workload as a {\em histogram}  -- {\em a distribution of query templates}. A histogram-based representation allows capturing the underlying statistical distribution of the queries by grouping them into bins or templates. Histograms have been used in different domains to aggregate multiple observations and obtain approximate data distributions \cite{Koloniari}. To simplify the experiment setup, the current design of LearnedWMP uses fixed-length workloads. However, the design can easily be extended to work with variable-length workloads. In {\em the final step}, using training workloads and their historical collective actual memory usage, we train a regression model that learns to estimate the working memory usage of an unseen workload based on the distribution of query templates. As the model learns from diverse training workloads, its accuracy at estimating the working memory of workloads will improve with time.

\stitle{Contributions}. We summarize our key contributions as follows: 
\begin{itemize}
\item We propose LearnedWMP, a novel prediction model that can estimate the working memory demand of a batch of SQL queries in a workload at once. This is a departure from the state of the practice and the state-of-the-art methods, which estimate the memory demand of each query separately. To the best of our knowledge, this is the first attempt to predict memory demand at the workload level using machine learning (ML) techniques. 
    \item We formulate the problem of workload memory prediction as {\em a distribution regression problem}, which learns a regressor function from workloads represented as distributions of query templates. We use ML to learn the regressor without relying on hand-crafted query-level or operator-level features. 
    \item We devise unsupervised ML methods to group queries of similar memory needs to reduce the computational overhead of workload memory usage estimation significantly.
    \item We extensively evaluate our LearnedWMP model employing three database benchmarks. These experimental results demonstrate the merit of our proposed technique in workload-based query processing and resource estimation.
    
\end{itemize}

\stitle{Summary of Experimental Results.} 
Our extensive experiments demonstrate that LearnedWMP substantially decreased memory estimation errors when compared to the current state-of-the-art practices, achieving an impressive improvement of 47.6\%. Our experiments are conducted over transactional and analytical database benchmarks that are used to train and evaluate our model. LearnedWMP accepts as input a workload and returns the workload’s estimated working memory demand for the workload. 
Our findings indicate that, during training and inferencing, the LearnedWMP model and its variant models were 3x to 10x faster compared to alternative ML models. Also, LearnedWMP-based models were at least 50\% smaller in most cases. Furthermore, our study delved into the performance and impact assessment of different components and parameters within the model. These included various learning models, workload sizes, numbers of query templates, and techniques for learning query templates (such as query plan-based and rule-based approaches). 

\stitle{DBMS Integration \& Broader Impact}. A DBMS vendor can pre-train a LearnedWMP model using sample training workloads and ship the model into the DBMS product. With the deployment of the DBMS on an operational site, the pre-trained LearnedWMP model can immediately generate memory estimations, which may not be highly accurate initially. DBMS can collect training workloads from its operational environment and use them to re-train the LearnedWMP model to improve the model's estimation accuracy. Major DBMS products now include ML infrastructure, which they can use for hosting, training, and serving the LearnedWMP model in the database \cite{IBMwebpage, MSwebpage,oracelwebpage}.

\smallskip\noindent\textbf{Paper Organization}. The rest of the paper is organized as follows. Section \ref{sec:problem} introduces key terminology, notations, and the problem formally. Section \ref{sec:methodology} provides an overview of LearnedWMP, including details of the training and inference stages. Section \ref{sec:evaluation} presents an experimental evaluation of LearnedWMP. Section \ref{sec:related} reviews the related work. Section \ref{sec:conclusions} concludes the paper.

\section{Preliminaries and the Problem}
\label{sec:problem}

In this section, we introduce notation and preliminaries to assist in defining the {\em workload memory prediction problem}. Next, we formally present our novel approach to representing and solving the problem as {\em a distribution regression problem}.



\stitle{Query}
Let $q$ = $(e, p, m)$ be a single SQL query where ({\em i}) $e$ is a query expression received from a database user, ({\em ii}) $p$ is a query execution plan generated by the DBMS optimizer for evaluating $e$, and ({\em iii}) $m$ is the actual highest working memory usage of the query for $p$. $m$ is available only for training queries that the DBMS has already executed. For unseen queries, $m$ is unknown.

\stitle{Workload}
Let $w$ = $(\mathcal{Q}, y)$ be a workload, which consists of ({\em i}) $\mathcal{Q}$, a set of queries where $q_i \in \mathcal{Q}$ is a tuple $(e_i, p_i, m_i)$, as per def. 2.1, and ({\em ii}) $y$ is the sum of the actual highest working memory utilization of all queries in $\mathcal{Q}$ after the DBMS executes them.
\begin{equation}
\label{eq:actual-memory}
    y = \max_{i=1}^{|\mathcal{Q}|} m_i 
\end{equation}
    
\noindent $y$ value of a workload (eq. 1 \ref{eq:actual-memory}) is only present for the training workloads executed by the DBMS. In the inference phase, LearnedWMP receives only Q, a collection of queries without y.

\stitle{Workload Memory Prediction}
Let us assume a training corpus of $n$ workloads as follows: 
 \begin{align}
 \{(w_1,y_1),\ldots,(w_n,y_n)\} 
  \label{eq:workload-history}
 \end{align}
Here, each tuple, $(w_i,y_i)$ corresponds to the highest historical working memory utilization $y_i$ of all queries in the workload $w_i$. Now, given an unseen workload $w$, we wish to learn a predictor function $\hat{f}(\cdot)$ that can accurately estimate the workload $w$'s highest working memory usage $y$:
\begin{equation}
    \hat{f}(w) = y
\end{equation}

\stitle{Working Memory} is a memory region used to store intermediate results, temporary data structures, and execution context information while executing database operations such as sorting and aggregation. A query may use the working memory up to a limit that is controlled by a DBMS setting. Working memory size in a DBMS varies based on system configuration, workload, and available resources. In the rest of this paper, for brevity, we may refer to working memory as only memory. 

\vspace*{5pt}

\noindent We formulate estimating memory usage of an unseen workload as a distribution regression problem \cite{Gretton,poczos2013distribution}, where the estimate is computed from an input probability distribution - the distribution of queries $\mathcal{Q}$ among templates $\mathcal{T}$. 

\stitle{Query Templates} Let $\mathcal{T}$ = \{$t_1$,...,$t_k$\} be a set of $k$ query templates. A query template $t_i \in \mathcal{T}$ represents a class of queries with similar plan characteristics and memory requirements. Any query $q$ can be mapped to a query template $t_i \in \mathcal{T}$.

\stitle{Workload Histogram}
Let $w$ be a workload consisting of a set of $\mathcal{Q}$ queries. $c_i$  is the number of queries in $\mathcal{Q}$ that can be mapped to query template $t_i \in \mathcal{T}$. 
The counts of queries in $\mathcal{Q}$ that map to different query templates in $\mathcal{T}$ are recorded in a 1-$d$ vector of length $k=|\mathcal{T}|$. We call this vector a {\em workload histogram} $\mathcal{H}$.   Here,
$\mathcal{H} = [c_1, ...,c_k]$ and 
\begin{equation}
\sum_{i=1}^{k} c_i = |\mathcal{Q}|    
\end{equation}
\noindent From such an input distribution, encoded in a {\em workload histogram}, a distribution regression function computes as estimated memory usage for the workload. Assume we have a training corpus of $n$ workload histograms, one histogram per workload, as follows: 
 \begin{align}
 \{(\mathcal{H}_1,y_1),\ldots,(\mathcal{H}_n,y_n)\} 
  \label{eq:histogram-history}
 \end{align}
Here, each tuple, $(\mathcal{H}_i,y_i)$, corresponds to a single workload; $\mathcal{H}_i$ is the workload histogram, and $y_i$ is the collective historical memory utilization of all queries in the workload.
%
%
On the workload histogram, we assume the following:
\begin{enumerate}
    \item The distribution of queries among the query templates (i.e., the workload histogram bins) is uniform.
    \item Query templates are independently and identically distributed.
    \item An underlying function, $f(\cdot)$, exists that can accurately compute any workload's memory usage, $y$, from the workload histogram, $\mathcal{H}$. 
    \begin{equation}
        f(\mathcal{H}) = y
    \end{equation}
\end{enumerate}
\noindent We, however, neither know $f(\cdot)$ nor have access to the set of all possible workload examples to derive $f(\cdot)$.
We wish to learn a function, $\hat{f}(\cdot)$, an approximation of $f(\cdot)$, using the distribution of regression. From the input {\em workload histogram}, $\mathcal{H}$, of a workload, $\hat{f}(\cdot)$ can compute $\hat{y}$, an accurate estimate of the actual memory usage $y$. 
\begin{equation}
    \hat{f}(\mathcal{H}) = \hat{y}
\end{equation}
Using training workloads labeled with their actual memory usage, $\hat{f(\cdot)}$ learns to estimate the memory usage of unseen workloads. We expect that the larger and more diverse the training data set of workload examples is, the more precise the predictor $\hat{f(\cdot)}$ will be. Table \ref{table:notations} provides a summary of the key notations.



\begin{table}[t!]
    \caption{Summary of key notations}
    \centering
     \setlength{\tabcolsep}{1pt}
    \begin{tabular}{cp{0.84\linewidth}}
        \toprule[1.2pt]
         \textbf{Symbol} & \multicolumn{1}{c}{\textbf{Description}}  \\
         \midrule
         \textbf{$w$} & A workload.  \\
         \textbf{$\mathcal{Q}$} &  The set of queries in a workload.\\
         \textbf{$\mathcal{T}$} & The set of query templates in the DBMS. \\
         \textbf{$\mathcal{H}$} & $\mathcal{H} \in \mathbb{R}^k$ is a workload histogram, representing the distribution of queries in a workload $w$ over the $k$ query templates $\mathcal{T}$.\\
         \textbf{$\hat{f(\mathcal{H})}$} & The learned function (predictor) that predicts the memory demand of an input workload histogram $\mathcal{H}$.\\ 
         \textbf{$c_i$} & The number of queries in a workload $w$ that are mapped to a query template $t_i \in \mathcal{T}$.\\
         \textbf{$y$} & The actual collective historical memory utilization of all queries $\mathcal{Q}$ in a workload $w$.\\ 
         \textbf{$\hat{y}$} & The predicted collective memory demand of all queries $\mathcal{Q}$ in an unseen workload $w$ as estimated by $\hat{f(\cdot)}$.\\ 
         \bottomrule[1.2pt]
    \end{tabular}
    \label{table:notations}
\end{table}

\section{The L\MakeLowercase{earned}WMP Model}
\label{sec:methodology}


LearnedWMP comprises two stages: training and inference. The training stage employs an ML pipeline and dataset to build the model, while the inference stage utilizes the trained model to predict memory usage for unseen workloads. Fig. \ref{fig:ml-pipeline} offers an overview of the workflow, and we subsequently outline and delve into the technical details of each step.

\begin{figure*}
  \includegraphics[width=\textwidth]{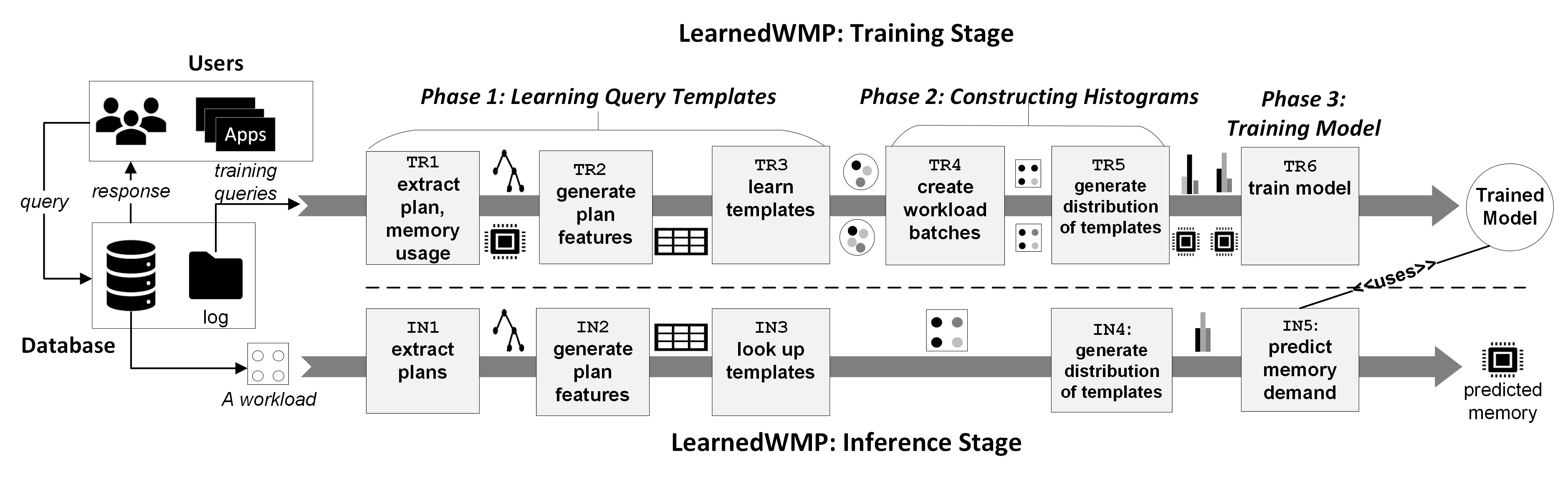}
  \caption{Overview of the \model model. Left: Users send queries to a DBMS; the DBMS processes and sends responses to the queries. Right-upper: the training steps of the \model model. Right-bottom: the steps of the inference stage.}
  \label{fig:ml-pipeline}
  \vspace{-10px}
\end{figure*}

\subsection{Overview of the \model 's pipeline}
\stitle{Users and the Database.} The top left section of Fig.~\ref{fig:ml-pipeline} illustrates user-database interaction, with the database interacting with two \model stages. The users and applications send SQL queries to the database. To calculate the memory utilization of a workload, we sum the highest memory usage for queries in that workload. The \model training pipeline (right of Fig.~\ref{fig:ml-pipeline}) periodically retrains the model using the latest query log dump.

\stitle{Training Stage.} In Fig. \ref{fig:ml-pipeline} (top), \texttt{TR1} through \texttt{TR6} are the steps of the training pipeline. Training begins with a set of training queries, $\mathcal{Q}_{train}$, collected from a dump of the DBMS query log. At \texttt{TR1}, from $\mathcal{Q}_{train}$, the pipeline extracts training queries, their final execution plans, and the actual highest working memory usage from the past execution. At \texttt{TR2}, from the query plans, the pipeline generates a set of $m$ features to represent the training queries as a $|\mathcal{Q}_{train}| \times m$ feature matrix. 
At \texttt{TR3}, the pipeline learns $\mathcal{T}$, a set of $k$ query templates, from the query feature matrix. the value of $k$ can be determined experimentally (cf. section \ref{sec:methodology-training}). At \texttt{TR4}, the pipeline equally divides the training queries of $\mathcal{Q}_{train}$ into a set of $n$ workloads, $\mathcal{W}$  = $|\mathcal{Q}_{train}|/s$. Each workload contains $s$, a constant, queries. We found a value of $s$ experimentally (cf. subsection \ref{sec:sensitivity}). At \texttt{TR5}, the pipeline generates a workload histogram $\mathcal{H}$ for each training workload $w$ = $(\mathcal{Q}, y)$ in $\mathcal{W}$.  $\mathcal{H}$ represents the distribution of queries of $\mathcal{Q}$ among $k$ query templates of $\mathcal{T}$. In addition, the collective actual highest memory utilization $y$ of the workload $w$ is computed by summing up the working memory utilization of each query $q_i \in \mathcal{Q}$. Each ($\mathcal{H}$, $y$) pair represents a supervised training example for training a regression model. At \texttt{TR6}, the model receives as input a collection of training examples of the form ($\mathcal{H}$, $y$). From these examples, the model learns a regression function $\hat{f}(\mathcal{H})$ to map an input histogram $\mathcal{H}$ to its working memory demand, $y$. At the end of \texttt{TR6}, the training pipeline produces a trained \model model.

\stitle{Inference Stage.} In Fig.~\ref{fig:ml-pipeline} (bottom), \texttt{IN1} through \texttt{IN5} are the steps of the  \model inference pipeline, which generates estimated memory usage of an unseen workload $w$, consisting of $\mathcal{Q}$ queries. Step \texttt{IN1} collects the query plans of the queries $\mathcal{Q}$ in $w$; step \texttt{IN2} generates the feature vectors for these plans. Step \texttt{IN3} assigns each query $q_i \in \mathcal{Q}$ to a template $t_i \in \mathcal{T}$, from which \texttt{IN4} constructs a workload histogram $\mathcal{H}$. The final step, \texttt{IN5}, uses the histogram $\mathcal{H}$ as input to the \model model and predicts the memory usage $\hat{y}$ of the workload, $w$.

\subsection{L\MakeLowercase{earned}WMP: Training Stage}
\label{sec:methodology-training}

The \model model is trained in six steps (\texttt{TR1} - \texttt{TR6}), which we can semantically group in three phases (see Fig.~\ref{fig:ml-pipeline}). The first phase uses historical queries from a DBMS's query log to learn a set of {\em query templates}. The second phase prepares the {\em training dataset} for the learning algorithm. The third phase uses the training dataset to train a {\em regression model}, which can predict the memory usage of an input workload.
\subsubsection{Phase 1: Learning Query Templates} 
\label{subsection:learning-query-templates}
We assign queries to templates based on the intuition that queries with similar query-plan characteristics and cardinality estimates have similar execution-time memory usage. In each workload, by grouping similar queries in the same templates or clusters, we expect to model the memory requirements of the queries more efficiently and speed up the computation of training and inference stages. We are not looking for an optimal template assignment for each query, which will require computation of operator-level features, increase computation overhead, and outweigh the acceleration we hope to gain from compressing queries into templates. Also, estimating the cost of individual queries is a separate research problem \cite{ganapathi2009predicting, li2012robust} that we are not addressing in our current research. Instead, we rely on a best-effort algorithmic principle for assigning each query to a template. The template assignment needs to be just good enough, not highly accurate. This allows for designing simple but efficient methods for assigning queries to templates that neither jeopardize the runtime cost nor the ML approach. This is not an algorithmic simplification but rather an algorithmic design choice.

\noindent To assign queries to templates we use standard $k$-means clustering algorithm \cite{macqueen1967classification}. Algorithm \ref{algo:kmeans} describes the steps of \textsc{GetTemplates} function that uses training queries of $\mathcal{Q}_{train}$ to learn a set of $k$ query templates, $\mathcal{T}$. For each query in $\mathcal{Q}_{train}$, \textsc{GetTemplates()} first extracts the query plan (line 5). A query plan is a tree-like structure where each node corresponds to a database operator. The plan's execution begins at the leaf nodes and completes at the root node. The output of the root is the result of executing the entire query. Each node has an input and an output. When applicable, a node includes statistics, such as estimated pre-cardinality and post-cardinality of the operator. For each operator type in a query plan, \textsc{GetTemplates()} counts its frequency and aggregate cardinalities of all its instances. The frequency count and aggregated cardinality of each operator type are retrieved (line 6) and used to represent the {\em query features}. Finally, the $k$-means algorithm \cite{macqueen1967classification} uses these query feature vectors to learn $k$ clusters, each one representing a query template $t_i \in \mathcal{T}$ (line 8). We use the elbow method \cite{raschka2019python} to tune the value of $k$. Fig.~\ref{fig:feature-extraction} provides an illustrative example query, where its associated query plan tree (below) has five unique operators: \texttt{TBSCAN}, \texttt{HSJOIN}, \texttt{INDEX SCAN}, \texttt{SORT}, and \texttt{GROUP BY}. Since each operator type provides a (count, cardinality) pair of features (\# of operators, total cardinality), this sample query plan has 10 features - 2 features for each of the 5 operators. \textsc{GetTemplates()} encodes these features in a 1-$d$ vector as follows: 
\noindent [4, 139532.48, 3, 50224.6, 1, 3201, 1, 134179, 1, 48873.6]. We borrowed this approach to featurizing queries from \cite{ganapathi2009predicting}, who found this set of features to be a good choice for predicting query performance. 
\begin{algorithm}[H]
 \caption{Learning query templates with $k$-means clustering}
 \label{algo:kmeans}
    \begin{algorithmic}[1]
  \State $\mathcal{Q} {train} \leftarrow \{q_1,\ q_2,...,\ q_n\}$ \Comment{$\mathcal{Q}_{train}$ is a set of historical training queries collected from a DBMS.} 
     \Function{GetTemplates}{$\mathcal{Q}_{train}$}
          \State $Array\ Z \leftarrow [][]$ \Comment{feature matrix for $\mathcal{Q}_{train}$}
         \For{$q_i \in\ \mathcal{Q}_{train}$}
         \State $plan_i$ = getQueryPlan($q_i$)
         \State $features_i$ = getFeatures($plan_i$)
        \State Z.insert($features_i$)
         \EndFor
        \State $\mathcal{T} \leftarrow kmeans(Z, k)$ \Comment{learns templates using kmeans}
        \Return $\mathcal{T}$ \Comment{\textit{k} learned query templates}
    \EndFunction
 \end{algorithmic}
\end{algorithm}

\subsubsection{Phase 2: Constructing Histograms from Workloads}
In this phase, \model performs two tasks: ({\em i}) partitions training queries of $\mathcal{Q}_{train}$ into a set workloads, $\mathcal{W}$ and ({\em ii}) from each training workload, $w_i \in \mathcal{W}$, constructs a histogram that represents the distribution of its queries among the query templates set $\mathcal{T}$. Histograms have played a significant role in estimating query plans cost \cite{bruno2002exploiting}. \model randomly divides training queries from $\mathcal{Q}_{train}$ into $m$ training workloads, where $m = |\mathcal{Q}_{train}| / s$ with $s$ being a constant number of training queries per workload. The value for $s$ depends on the application domain and can be empirically found (cf. subsection \ref{sec:sensitivity}). As defined in \textit{definition 2.2}, each training workload, $w_i$, is a tuple $(\mathcal{Q}, y)$, where $\mathcal{Q}$ is a collection of queries and $y$ is their collective memory usage from the past execution.\\
Algorithm \ref{algo-kmeans-Templatizing} describes the steps of phase 2. It takes as input a training workload, $w$, and a set of query templates, $\mathcal{T}$, which were learned in phase 1. In \texttt{lines 6-7}, for each query $q_i \in \mathcal{Q}$, the algorithm extracts the query execution plan and the features from the plan. Since phase 1 already computed these features, \texttt{line 7} reuses the values from the previous computation. Using these features, \texttt{line 8} looks up the query template, $t_j \in \mathcal{T}$, for $q_i$. After assigning each query, $q_i \in \mathcal{Q}$, to a template, the algorithm counts the number of queries in $\mathcal{Q}$ in each template, $t_j \in \mathcal{T}$ (\texttt{line 10}) and stores the counts in a histogram, a 1-d count vector, $\mathcal{H} = [c_1, ...,c_{k=|\mathcal{T}|}]$. The length of $\mathcal{H}$ is $k$, corresponding to the number of query templates in $\mathcal{T}$. Each count $c_j \in \mathcal{H}$ is the number of queries in $w$ that are associated with the query template $t_j \in \mathcal{T}$. These counts add up to $s=|\mathcal{Q}|$, the workload batch size:
\begin{equation}
\sum_{j=1}^{k=|\mathcal{T}|} c_j = s   
\end{equation}

\noindent The histogram $\mathcal{H}$ is the distribution of queries in workload $w$ among the query templates set $\mathcal{T}$. The histogram will be sparse with many zeros as a workload is not expected to contain queries that belong to each query template of $\mathcal{T}$. At the final step (\texttt{line 11)}, the algorithm returns a pair ($\mathcal{H},y$), where $y$ is the collective memory usage of all $q \in \mathcal{Q}$. ($\mathcal{H},y$) becomes a labeled example for training a supervised ML model in Phase 3.

\begin{figure}[t]
    \includegraphics[width=0.45\textwidth]{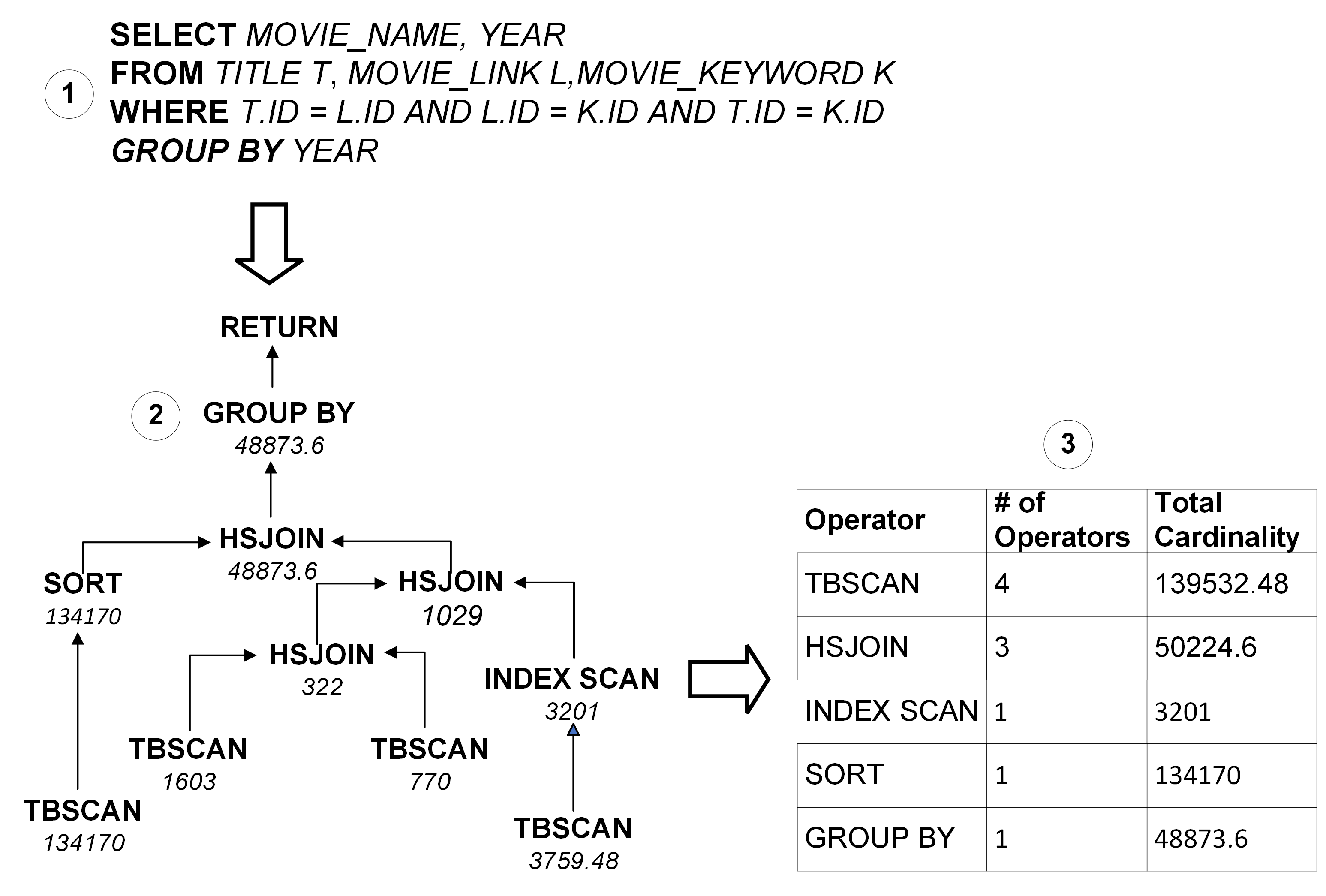}
  \caption{Extracting query features from a query. An example query (top) is executed by the query plan (left), leading to the extraction of the query features (right). The query features are used to learn a set of query templates $\mathcal{T}$, of size $k$.}
  \label{fig:feature-extraction}
  \vspace{-10px}
\end{figure}

\noindent Fig.~\ref{fig:binning} shows an example of constructing a histogram $\mathcal{H}$ of 4 bins, $k=|\mathcal{T}|=4$. The example uses an input workload, $w$, with 9 queries, $s=|\mathcal{Q}|=9$. $3$ of $4$ histogram bins are populated with nonzero values. The remaining bin has a zero value as its corresponding query template has no queries in $w$. The histogram vector is \texttt{[3, 4, 0, 2]}. The value of $y$ is the total actual memory usage of the historical memory usage of all 9 queries in $w$. Let's assume $y$ = \textit{125} MB. The output pair for this example workload is (\texttt{[3, 4, 0, 2]}, \textit{125}). 

\subsubsection{Phase 3: Training a Distribution Regression Deep Learning Model}
In this phase, we train a regression model for predicting workload memory usage. The trained model takes an input workload, represented as a histogram of query templates, and computes the workload's memory usage. We explored several ML and deep learning (DL) techniques to train the model. In this subsection, we will present the design and implementation of a DL model for our regression model. Section \ref{sec:other-ml-methods} describes the other ML algorithms we explored for the model training. DL recently had several algorithmic breakthroughs and has been highly successful with many learning tasks over unstructured data. For example, DL models for image recognition and language translation are now highly accurate \cite{shwartz2021tabular}. Additionally, DL can be useful in learning a non-linear mapping function between input and output without requiring low-level feature engineering. In our case, we have dual complexities: the input is a complex distribution of query templates, and there is a complex relationship between the distribution of query templates and its collective memory demand. We wanted to explore the effectiveness of deep learning for the problem. 

\noindent\textbf{Multilayer Perceptron (MLP) Model}. Various deep learning networks have been developed for unstructured data, such as convolution neural network (CNN) \cite{lecun1989backpropagation} for images; recurrent neural network (RNN) \cite{salehinejad2017recent} and transformers \cite{vaswani2017attention} for sequential data, and graph neural networks (GNN) for graphs \cite{murphy2022probabilistic}. Unstructured data often has variable-length input vectors with individual elements lacking meaning in isolation \cite{murphy2022probabilistic}. However, in our case, the input vector for each workload is structured and has a fixed length corresponding to the number of query templates. Each vector element represents the workload queries of a specific template. 
Since we want to learn a regression function from fixed-length input vectors, the multilayer perceptron (MLP) is a suitable choice for learning a regression function from fixed-length input vectors due to its assumption of fixed input dimension and flexibility in architecture~\cite{murphy2022probabilistic, subasi2007eeg}. 
MLP architecture consists of input, hidden, and output layers, and the optimal number of hidden layer neurons is determined through trial and error since there is no analytical method to precisely determine the ideal number crucial to prevent overfitting and ensure generalization \cite{haykin2009neural}.
In our case, from $n$ training examples $(\mathcal{H}_1, y_1), (\mathcal{H}_2, y_2),\ldots,(\mathcal{H}_n, y_n)$, a MLP model learns a function $f(\cdot): R^k \rightarrow R^o$, where $k$ is the number of dimensions for input $\mathcal{H} = [c_1, ...,c_{k=|\mathcal{T}|}]$ and $R^o$ is the scalar output $y$.

\begin{algorithm}[t]  
    \begin{algorithmic}[1]
    \State \ $w \leftarrow (\mathcal{Q}, y)$ \algorithmiccomment{A training workload}
    \State $\mathcal{T} \leftarrow \{t_1,...,\ t_{k}\}$  \algorithmiccomment{A set of $k$ query templates}
    \Function{BinWorkload}{$w$, $\mathcal{T}$}
        \State $Array\ \mathcal{H}[0\ ...\ (k-1)] \leftarrow 0$
        \For{$q_i\ in\ \mathcal{Q}$}
        \State $plan_i$ = getQueryPlan($q_i$)
        \State $features_i$ = getFeatures($plan_i$)
        \State $q_i.template$ = findTemplate($features_i$)
        \EndFor
        \For{$t_j\ in\ \mathcal{T}$}
        \State $\mathcal{H}[j]$ = $countTemplateInstances(t_j, w)$
        \EndFor
        \State \Return $(\mathcal{H}, y)$
    \EndFunction
  \end{algorithmic}
  \caption{Histogram construction from training workloads}
  \label{algo-kmeans-Templatizing}
\end{algorithm}


\noindent\textbf{Activation Function}. 
The activation function in each hidden layer determines how the input is transformed. We explored two options: linear and Rectified Linear Unit (ReLU). For simpler datasets with fewer query templates, linear activation performed better, while ReLU proved more effective for complex datasets with more query templates. ReLU enables improved optimization with Stochastic Gradient Descent and efficient computation and is scale-invariant.

\noindent\textbf{Loss Function}. Depending on the problem type, an MLP uses different loss functions. For the regression task, we use the mean squared error loss function as follows:
\begin{equation} 
Loss(\hat{y},y,W) = \frac{1}{2N}\sum_{i=1}^N||\hat{y}_i - y_i ||_2^2 + \frac{\alpha}{2N} ||W||_2^2
\end{equation}
where $y_i$ is the target value; $\hat{y}_i$  is the estimated value produced by the MLP model; $\alpha ||W||_2^2$ is an L2-regularization term (i.e., penalty) that penalizes complex models; and $\alpha > 0$ is a non-negative hyperparameter that controls the magnitude of the penalty.
The MLP begins with random weights and iteratively updates them to minimize the loss by propagating the loss backward from the output layer to the preceding layers and updating the weights in each layer. The training uses stochastic gradient descent (SGD), where the gradient $\nabla Loss_{W}$ of the loss with respect to the weights is computed and subtracted from $W$. More formally, 
\begin{equation}
W^{i+1} = W^i - \epsilon \nabla {Loss}_{W}^{i}
\end{equation} where $i$ is the iteration step, and $\epsilon$ is the learning rate with a value larger than 0. The algorithm stops after completing a preset number of iterations or when the loss doesn't improve beyond a threshold.

\begin{figure}[t]
  \includegraphics[width=0.45\textwidth]{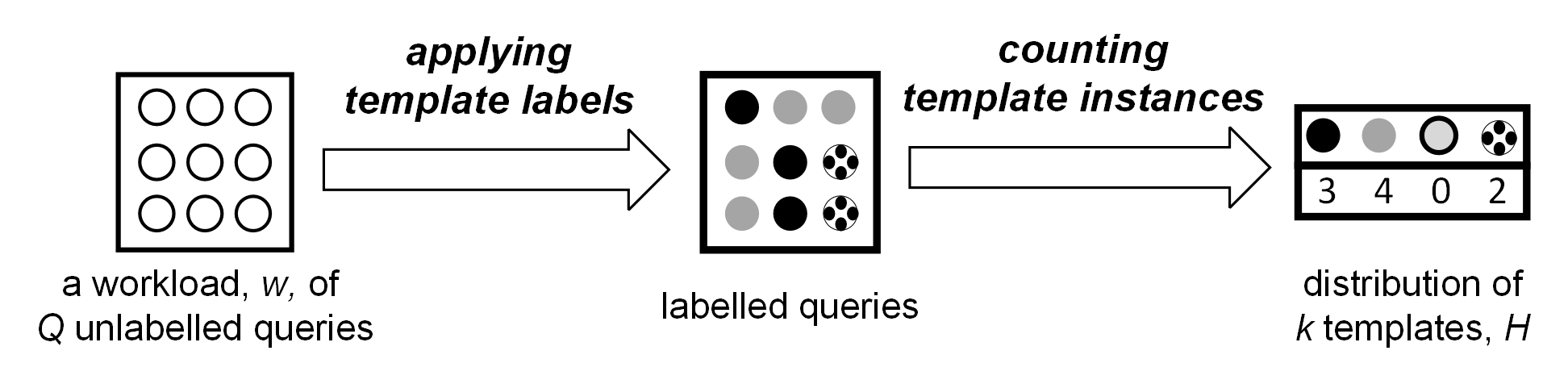}
  \caption{An example of generating a histogram $\mathcal{H}$ from a training workload $w=(\mathcal{Q}, y)$, where $|\mathcal{Q}|=9$ and $k=4$.}
  \label{fig:binning}
  \vspace{-10px}
\end{figure}

\noindent\textbf{Optimizer}. We compared L-BFGS \cite{liu1989limited} and Adam \cite{kingma2014adam} optimizers using two datasets --- a small dataset and a relatively large one. 
For the small dataset, L-BFGS was more effective than Adam as it ran faster and learned better model coefficients. In contrast, Adam worked better with the large dataset. Our observation is consistent with {\em scikit-learn}'s \texttt{MLPRegressor}\cite{regressorSklearn}.

\smallskip\noindent\textbf{Hyperparameter Tuning of the MLP Model}. 
We tuned hyperparameters, including the number of hidden layers, the number of nodes in each layer, the optimizer, and the dropout rate. Given the large dataset and parameter search space, we employed randomized search using \texttt{scikit-learn} library \cite{SklearnSearch}. We found a neural network architecture with eight layers (input, six hidden, and output), and the hidden layers contained 48, 39, 27, 16, 7, and 5 nodes from left to right, while the input layer received workload histograms or query distribution, and the output layer provided estimated memory demand.

\smallskip\noindent\textbf{Model Complexity}. Assume $n$ training samples, $k$ features, $l$ hidden layers, each containing $h$ neurons --- for simplicity, and $o$ output neurons. The time complexity of backpropagation is $O (n \cdot k \cdot h^l \cdot o \cdot i)$, where $i$ is the number of iterations. 
\subsubsection{Other machine learning methods} 
\label{sec:other-ml-methods}
Besides deep learning networks, for a comparative analysis, we explored four additional ML techniques to train \model models. They include a {\em linear} and three {\em tree-based} techniques. For the linear model, we picked \textbf{Ridge}, a popular method for learning regularized linear regression models \cite{raschka2019python}, which can help reduce the overfitting of the linear regression models. From the tree-based approaches, we used \textbf{Decision Tree (DT)}, \textbf{Random Forest (RF)}, and \textbf{XGBoost (XGB)}. 
DT uses a single tree for predictions, while Random Forest employs an ensemble of trees that consider random feature subsets, resulting in better generalization and outlier handling capabilities \cite{raschka2019python}. 
Our final tree-based model was XGBoost \cite{chen2016xgboost}, a gradient boosting tree technique that has achieved state-of-the-art performance for many ML tasks based on tabular data \cite{shwartz2021tabular}. 

\subsection{L\MakeLowercase{earned}WMP: Inference Stage}
\label{sec:methodology-inference}

Algorithm \ref{alg:inference} describes the steps of \textsc{PredictMemory()} function that estimates the memory demand of an unseen workload, $w$. The function operates with two models: a trained $k$-means clustering model, $\mathcal{T}$, which represents a set of learned query templates, and a trained predictive model for estimating workload memory demand. \textsc{PredictMemory()} receives as input an unseen workload whose memory demand needs to be estimated. At \texttt{line 5}, the function generates a histogram vector, $\mathcal{H}$, a distribution of query templates. This step reuses the \textsc{BinWorkload()} function of algorithm \ref{algo-kmeans-Templatizing}. Next, \texttt{line 6} estimates working memory demand $\hat{y}$ of $w$ using $\mathcal{H}$.
\begin{algorithm}[t]
    \begin{algorithmic}[1]
    \State $\mathcal{T} \leftarrow \{t_1,...,\ t_{k}\}$  \algorithmiccomment{A set of $k$ query templates}
    \State $\hat{f}$ \algorithmiccomment{a workload memory estimation function}
    \State $w \leftarrow (\mathcal{Q})$ \algorithmiccomment{An unseen input workload}
    
    \Function{PredictMemory}{$w,\ \mathcal{T},\ \hat{f}$}
       \State $\mathcal{H}$ = $BinWorkload(w, \mathcal{T})$
        \State $\hat{y}$ = $\hat{f}(\mathcal{H})$
        \State \Return $\hat{y}$
    \EndFunction
    \end{algorithmic}
  \caption{Predict workload memory by rained MLP}
  \label{alg:inference}
\end{algorithm}

\section{Experimental Evaluation}
\label{sec:evaluation}
We would like to reiterate that LearnedWMP is an innovative model addressing the novel problem of predicting memory for a workload. In this section, we experimentally evaluate LearnedWMP through this prism. 
In our first set of experiments, we demonstrate how LearnedWMP estimates memory for a workload using different ML models and metrics, revealing a significant reduction in estimation error compared to alternative baselines for predicting query memory demand. Second, we show that LearnedWMP is more efficient in terms of training, inference time, and model size when compared to single-based query models. Finally, we conduct an analysis to study the impact of the major parameters of the LearnedWMP model and the choice of the template learning method.

\stitle{Baselines.} In the workload memory prediction problem, we aim to estimate the aggregate memory demand for a query batch (i.e., workload). As there are no existing libraries or methods for this novel problem, we develop and evaluate different variants of LearnedWMP and compare them with the state-of-the-art single-based models for predicting query memory demand and assessing their performance. 
\begin{itemize}
\item \textbf{LearnedWMP-based Methods.} LearnedWMP accepts as input a workload $w$ and returns the workload's estimated working memory demand, $\hat{y}$. As described in section \ref{sec:other-ml-methods}, besides the proposed MLP-based deep neural network (DNN) method, we can use other ML techniques to learn the memory estimation regression function. Thus, we explored additional ML techniques, such as Ridge, DT, RF, and XGB, in this experiment. We refer to the LearnedWMP-based models trained with different ML techniques as \textit{LearnedWMP-DNN}, \textit{LearnedWMP-Ridge}, \textit{LearnedWMP-DT}, \textit{LearnedWMP-RF}, and \textit{LearnedWMP-XGB}. 

\begin{figure*}[t!]
  \makebox[\linewidth][c]{%
     \centering
     \begin{subfigure}[b]{0.31\textwidth}
         \centering
         \includegraphics[width=\textwidth]{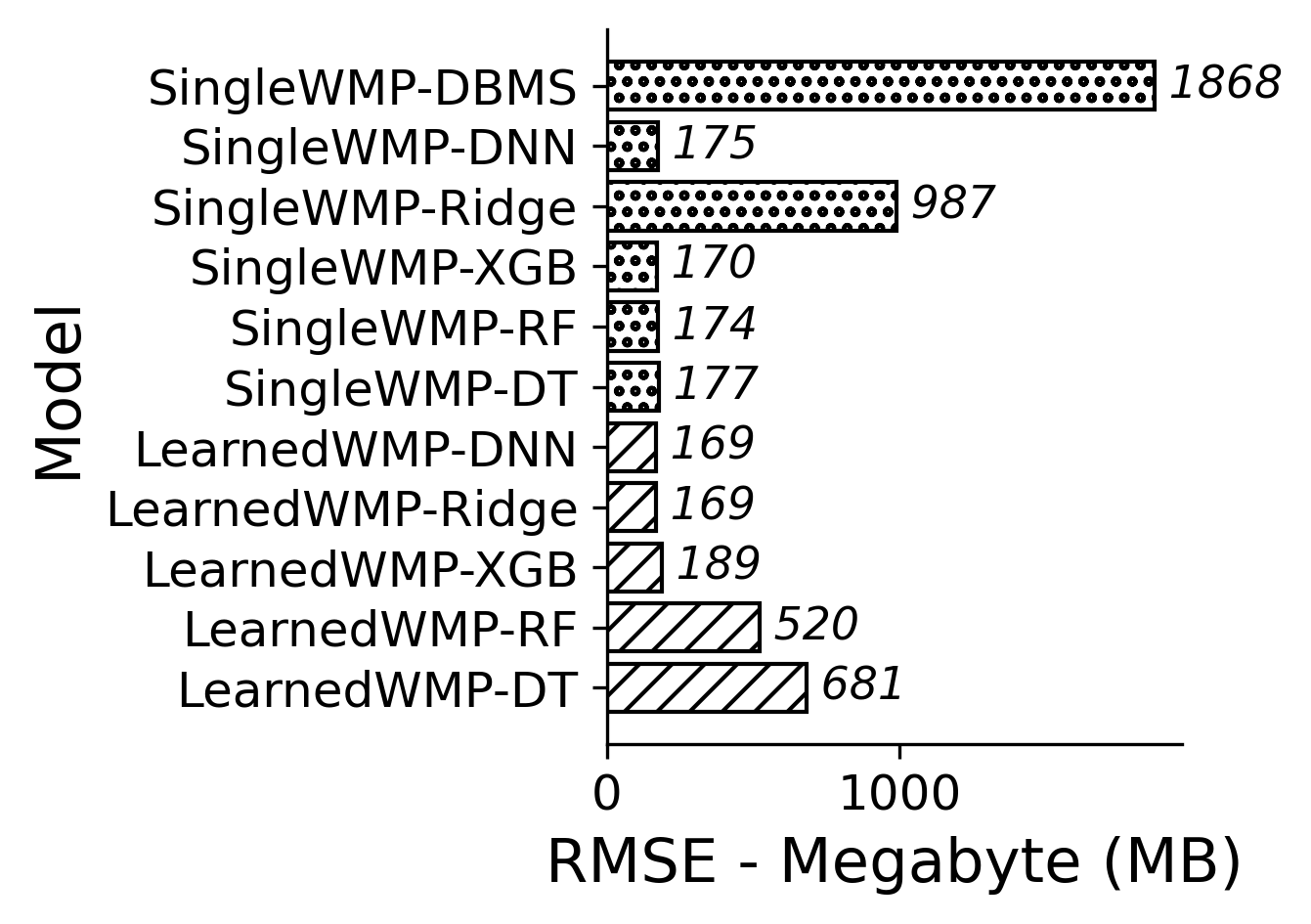}
         \caption{\texttt{TPC-DS}}
         \label{fig:TPC-DS-rmse}
     \end{subfigure}
    \hfill
     \begin{subfigure}[b]{0.31\textwidth}
         \centering
         \includegraphics[width=\textwidth]{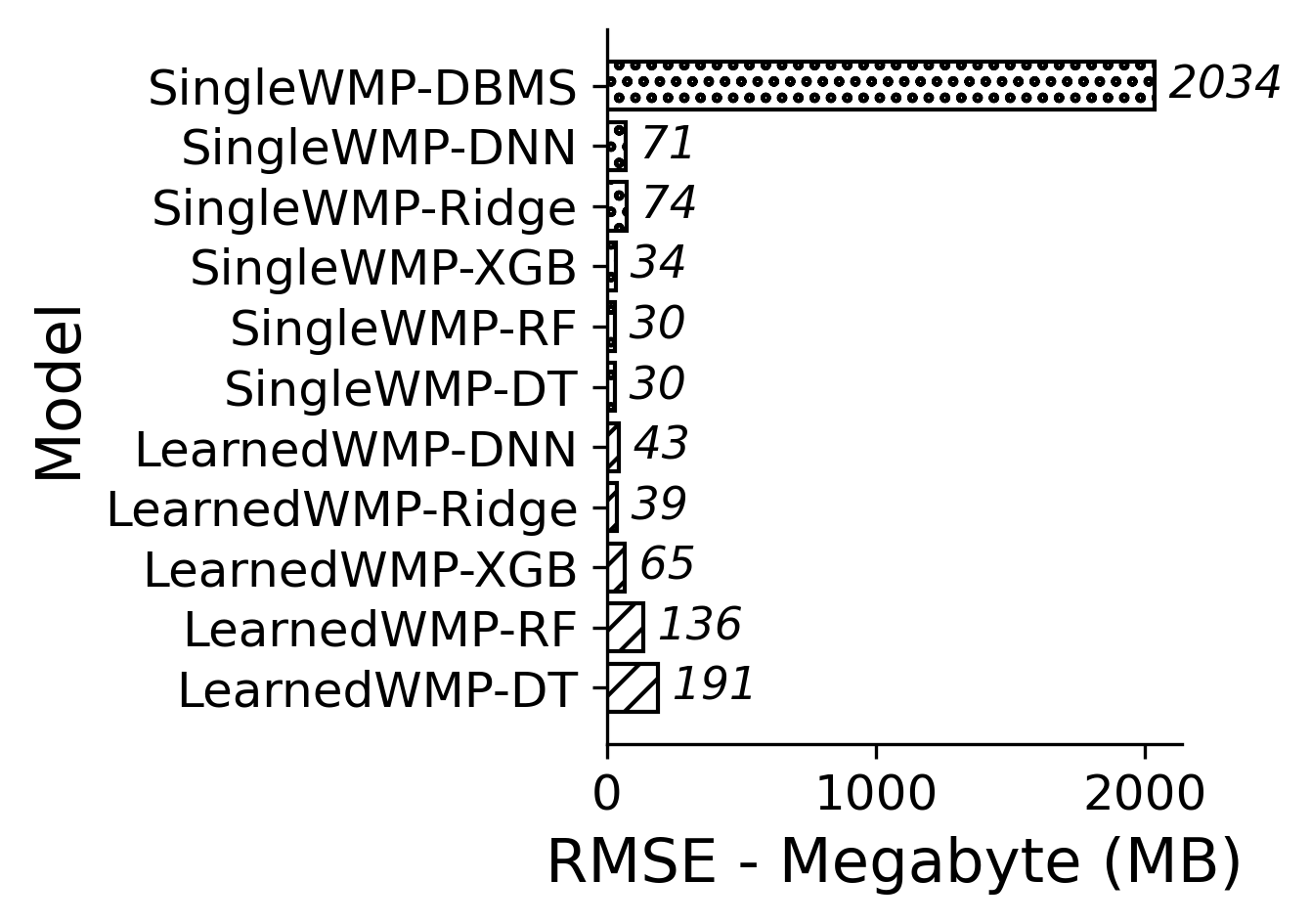}
         \caption{\texttt{JOB}}
         \label{fig:JOB-rmse}
     \end{subfigure}
     \hfill
     \begin{subfigure}[b]{0.31\textwidth}
         \centering
         \includegraphics[width=\textwidth]{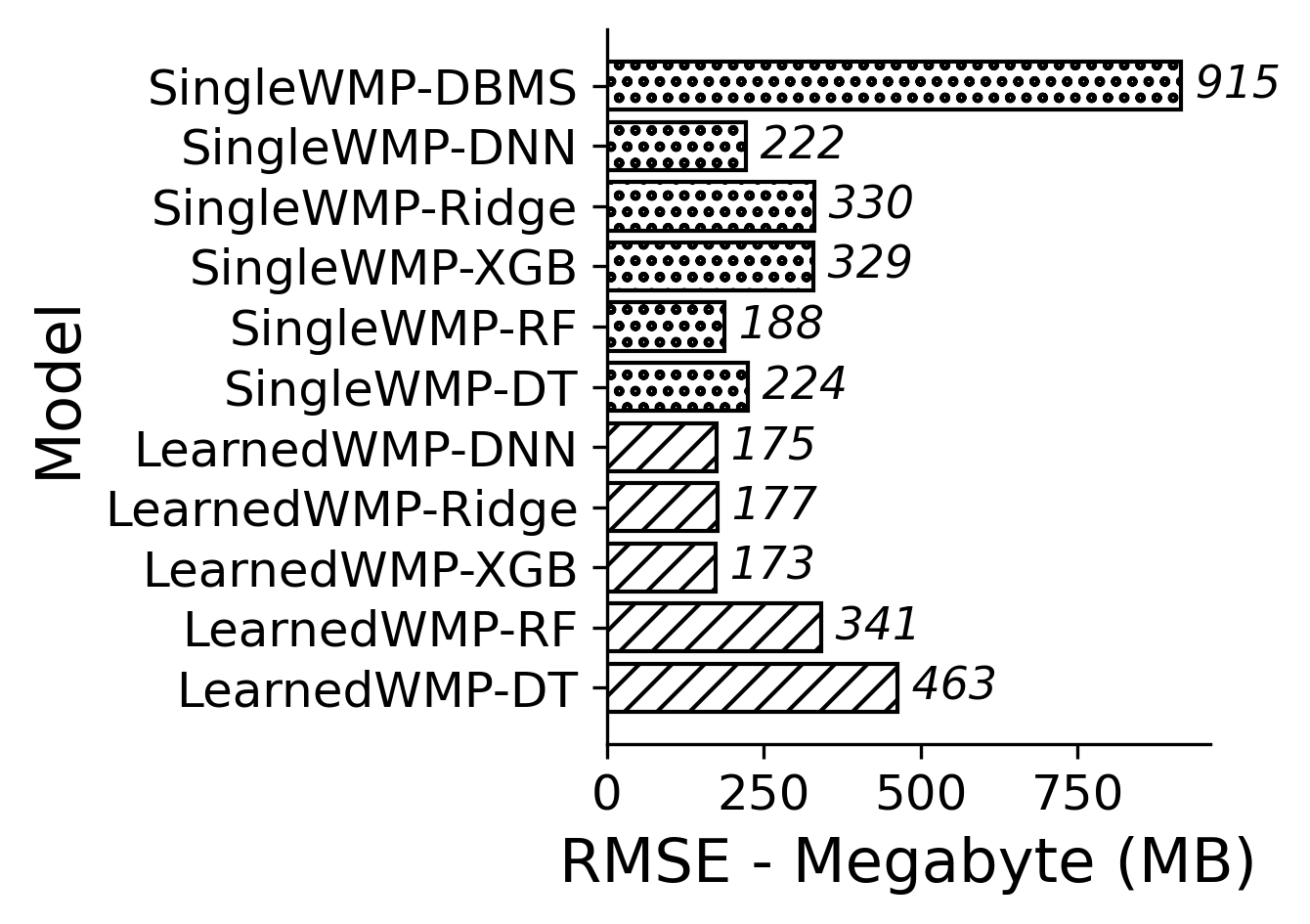}
         \caption{\texttt{TPC-C}}
         \label{fig:TPC-C-rmse}
     \end{subfigure}
     }
        \caption{Root Mean Squared Errors (smaller is better)}
        \label{fig:model-training-rmse}
        \vspace{-10px}
\end{figure*}

\item \textbf{SingleWMP-based Methods}. An alternative approach to estimate the working memory demand of a workload is to rely on single-query-based methods for memory prediction. In this approach, first, the highest working memory requirement for each query in the workload is estimated separately. Then, these individual estimates are summed up to produce the aggregate working memory estimation for the workload. We refer to this method as \underline{Single}-query based \underline{W}orkload \underline{M}emory \underline{P}rediction (SingleWMP). Formally, given a workload $w$, consisting of a set of $\mathcal{Q}$ queries, the estimated memory demand $\hat{y}$ of $w$ is:
\begin{equation}
\hat{y}=\sum_{i=1}^\mathcal{|Q|}{\hat{y}_{q_i}}
\end{equation}
where $\hat{y}_{q_i}$ is the estimated memory demand of a single query $q_i \in \mathcal{Q}$. In this approach, we use query plan features of each query as direct input to an ML algorithm. During the training phase, the algorithm additionally receives the historical memory usage of each training query. Using the pairs of query plan features and memory usage of many individual training queries, the algorithm learns a function to estimate the memory demand of individual queries. Unlike the LearnedWMP approach, the SingleWMP approach does not learn query templates from historical queries. Similar to LearnedWMP, we use DNN, Ridge, DT, RF, and XGB techniques to train five variations of the SingleWMP ML models. We refer to these variants of SingleWMP as \textit{SingleWMP-DNN}, \textit{SingleWMP-Ridge}, \textit{SingleWMP-DT}, \textit{SingleWMP-RF}, and \textit{SingleWMP-XGB}. An additional method of the SingleWMP approach is \textit{SingleWMP-DBMS}, which obtains the estimated memory usage for each query directly from a DBMS's query optimizer. SingleWMP-DBMS doesn't use ML in the memory estimation; instead, it relies on heuristics written by database experts. SingleWMP-DBMS represents {\em the current state of practice} in the commercial database management systems.
\end{itemize}

\stitle{Datasets.} We used three popular database benchmarks: TCP-DS\cite{nambiar2006making} 
Join Order Benchmark or JOB\cite{leis2015good}, and TPC-C\cite{leutenegger1993modeling}. TPC-DS and JOB are benchmarks for analytical workloads, whereas TPC-C is for transactional workloads. We used either its query generation toolkit or the seed query templates for each benchmark to generate queries for our experiment. We generated 93000 queries for TPC-DS, 2300 queries for JOB, and 3958 queries for TPC-C. For each benchmark, we randomly divided the queries into training ($\mathcal{Q}_{train}$) and test ($\mathcal{Q}_{test}$) partitions, with 80\% of the queries belonging to $\mathcal{Q}_{train}$ and the remaining 20\% to $\mathcal{Q}_{test}$. We grouped queries into workload batches from training and test partitions. We experimented with different batch sizes and found 10 to be a decent size to improve the memory estimation of our experimental workloads. We discuss our experiment with batch size parameter in section \ref{sec:sensitivity}.

\stitle{Evaluation metrics.} To evaluate the accuracy performance of various models, we use two accuracy metrics:
\begin{itemize}
    \item \emph{Root Mean Squared Error (RMSE)}: For measuring the accuracy performance of the LearnedWMP and SingleWMP models. We use $L_2$ loss or root mean squared error (RMSE), as follows:
    \begin{align}
    RMSE =  \sqrt{ \frac{\sum_{i=1}^N {(y_i-\hat{y_i})^2}}{N} }
    \label{eqmse}
    \end{align}
We seek to find an estimator that minimizes the RMSE.
\item \emph{IQR and Error Distribution}: While RMSE is convenient to use, it doesn't provide insights into the distribution of prediction errors of a model. Two models can have similar RMSE scores but different distributions of errors. For each benchmark, we compute the signed differences between the actual and the predicted memory estimates - the residuals of errors. We use the residuals to generate violin plots, which help us compare the interquartile ranges (IQR) and the error distributions of different models. IQR is defined as follows \cite{dekking2005modern}.
    \begin{equation}
    IQR = q_n(0.75) - q_n(0.25)
    \end{equation}
    Here, $q_n(0.75)$ is the 75th percentile or the upper quartile, and the $q_n(0.25)$ is the 25th percentile or the lower quartile. The range of values that fall between these two quartiles is called the interquartile range (IQR). IQR is shown as a thick line inside the violin in a violin plot. A white circle on the IQR represents the median. When a model's violin is closer to zero and has a smaller violin, it is more accurate. 
\end{itemize}
In addition to computing accuracy, for each ML-based LearnedWMP and SingleWMP model, we measured the {\em model size} in kilobyte ($kB$), the {\em training time} in millisecond ($ms$), and the inference time in microsecond ($\mu$).

\stitle{Experiments Design.} In Section\ref{sec:learnedwmp-accuracy}, we evaluate the performance of LearnedWMP-based models compared to that of SingleWMP-based models. The computational overhead of the LearnedWMP model is discussed in Section \ref{sec:trainingcost}. This includes the model size and runtime cost of training and inference of LearnedWMP-based models and how it compares to SingleWMP-based models'. Section~\ref{sec:sensitivity}
performs a sensitive study for the parameters and design choices of the LearnedWMP model. We conducted the experiments using a commercial DBMS instance running on a Linux system with 8 CPU cores, 32 GB of memory, and 500 GB of disk space.

\begin{figure*}[t!]
     \centering
     \makebox[\linewidth][c]{%
     \begin{subfigure}[b]{0.31\textwidth}
         \centering
         \includegraphics[width=\textwidth]{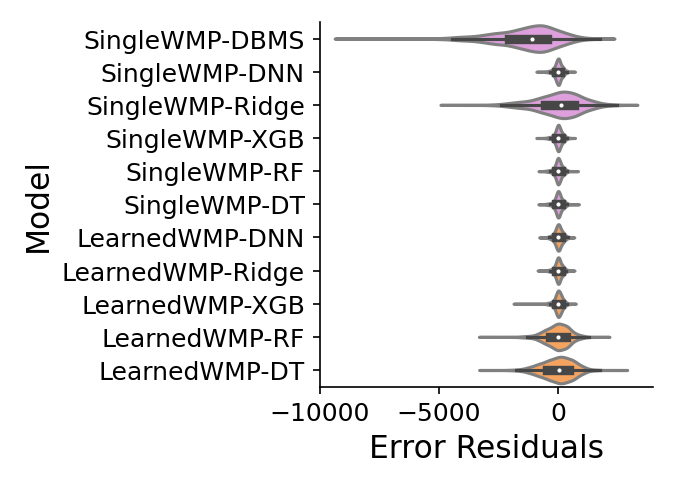}
         \caption{\texttt{TPC-DS}}
         \label{fig:TPCDS-mtl}
     \end{subfigure}
     \hfill
     \begin{subfigure}[b]{0.31\textwidth}
         \centering
         \includegraphics[width=\textwidth]{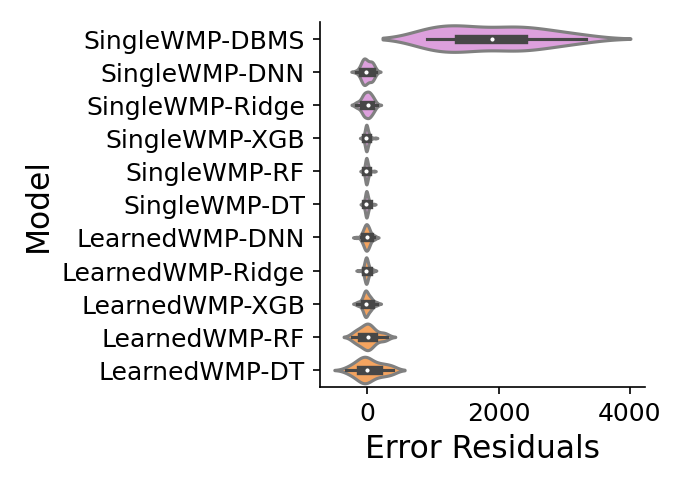}
         \caption{\texttt{JOB}}
         \label{fig:JOB-mtl}
     \end{subfigure}
    \hfill
     \begin{subfigure}[b]{0.31\textwidth}
         \centering
         \includegraphics[width=\textwidth]{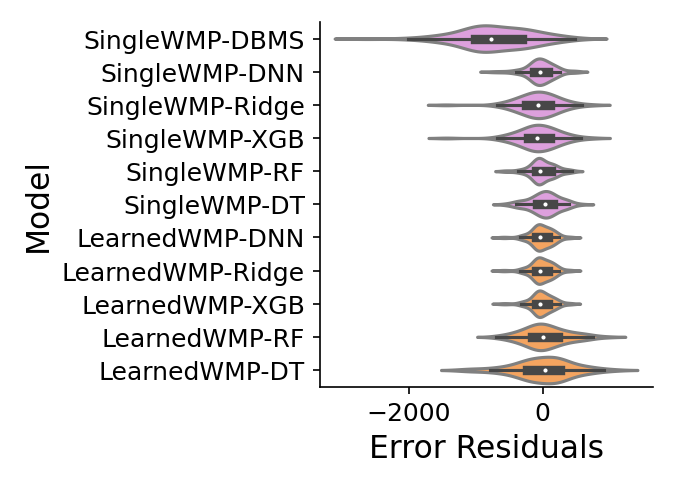}
         \caption{\texttt{TPC-C}}
         \label{fig:TPC-C-mtl}
     \end{subfigure}
     }
        \caption{Estimation Error Residuals Distributions}
        \label{fig:violin-plots}
        \vspace{-10px}
\end{figure*}

\subsection{LearnedWMP Accuracy Performance}
\label{sec:learnedwmp-accuracy}
We report on LearnedWMP's accuracy performance in terms of RMSE and the distribution of error residuals presented as violin plots for completeness.

\smallskip\noindent\textbf{RMSE}. Fig. \ref{fig:model-training-rmse} reports the RMSEs of SingleWMP-based and Learned\-WMP-based models. SingleWMP-DBMS represents the state of practice in commercial DBMSs. LearnedWMP models and ML-based SingleWMP models significantly outperformed the SingleWMP-DBMS model. For the TPC-DS, SingleWMP-DBMS's RMSE was 1868. In comparison, LearnedWMP-DNN and LearnedWMP-Ridge, the two best ML models, had an RMSE of 169, which was a 90.95\% 

 On RMSE, ML-based methods --- LearnedWMP-based models and SingleWMP-based ML models --- were significantly more accurate than SingleWMP-DBMS method. Using heuristics, SingleWMP-DBMS couldn't accurately capture the complex interactions between database operators within a query plan and produced large estimation errors. In contrast, using ML, LearnedWMP-based, and SingleWMP-based ML models learned to estimate the memory requirements of complex database workloads more accurately.

\smallskip\noindent\textbf{IQR and Error Distribution}. Fig. \ref{fig:violin-plots} compares the violin plots of different models. The violins of the SingleWMP-DBMS are wider and far from zero. DBMS's estimations are skewed towards either underestimation or overestimation and span a larger region. In contrast, ML-based estimates are balanced between over and under-estimations and are not skewed. The violins of ML-based models span a smaller range than SingleWMP-DBMS's. For TCP-DS, LearnedWMP-DNN's violin is centered at zero and small. For the same dataset, SingleWMP-DBMS's violin is skewed toward underestimation and larger when compared to the IQR of LearnedWMP-DNN.  We see a similar pattern when comparing the violins of SingleWMP-DBMS models with the violins of other ML models. 
Using human-crafted rules, SingleWMP-DBMS's estimation errors are not distributed between overestimations and underestimations. These static rules skew the estimations toward one direction. In contrast, ML-based models learn from real-world workloads --- which include examples of both overestimation and underestimation --- and learn to compute memory predictions that are not skewed in one direction. For instance, the memory estimation errors of the DNN and XGboost methods for both SingleWMP-based and LearnedWMP-based models are smaller and balanced.
 \vspace{-1px}
\subsection{Computational Overhead}
\label{sec:trainingcost}
 We evaluate the LearnedWMP-based and SingleWMP-based models overhead in terms of training time, inference time, and model size. All these metrics are important when considering embedding an ML model into the DBMS.

\stitle{Training and Inference Time.} Fig. \ref{fig:model-training-latency} reports the training time of all models\footnote{Note that for this set of experiments, we do not consider LearnedWMP-DBMS as it is not an ML model, and it doesn't have a training and an inference cost.}. For each benchmark dataset, LearnedWMP-based and SingleWMP-based methods use the same set of training queries as input. The singleWMP-based method uses individual training queries directly as input to the model. LearnedWMP batches the training queries into workloads, represents workloads as histograms, and uses the histograms as input to the models. Compared to SingleWMP-based models, the training of LearnedWMP-based models was significantly faster. For instance, with the TPC-DS dataset, SingleWMP-XGB was trained in 912.7 ms, whereas LearnedWMP-XGB was trained in 404 ms --- which is more than 2x faster. For all three benchmark datasets, we observe a similar trend: the training of a LearnedWMP-based model was faster than that of the equivalent SingleWMP-based model. The Ridge is the only algorithm that didn't demonstrate a significant difference in training time between the LearnedWMP and SingleWMP approaches. This is expected as Ridge is a linear algorithm that doesn't involve a sophisticated learning method.

\begin{figure*}[t!]
     \centering
     \makebox[\linewidth][c]{%
     \begin{subfigure}[b]{0.31\textwidth}
         \centering
         \includegraphics[width=\textwidth]{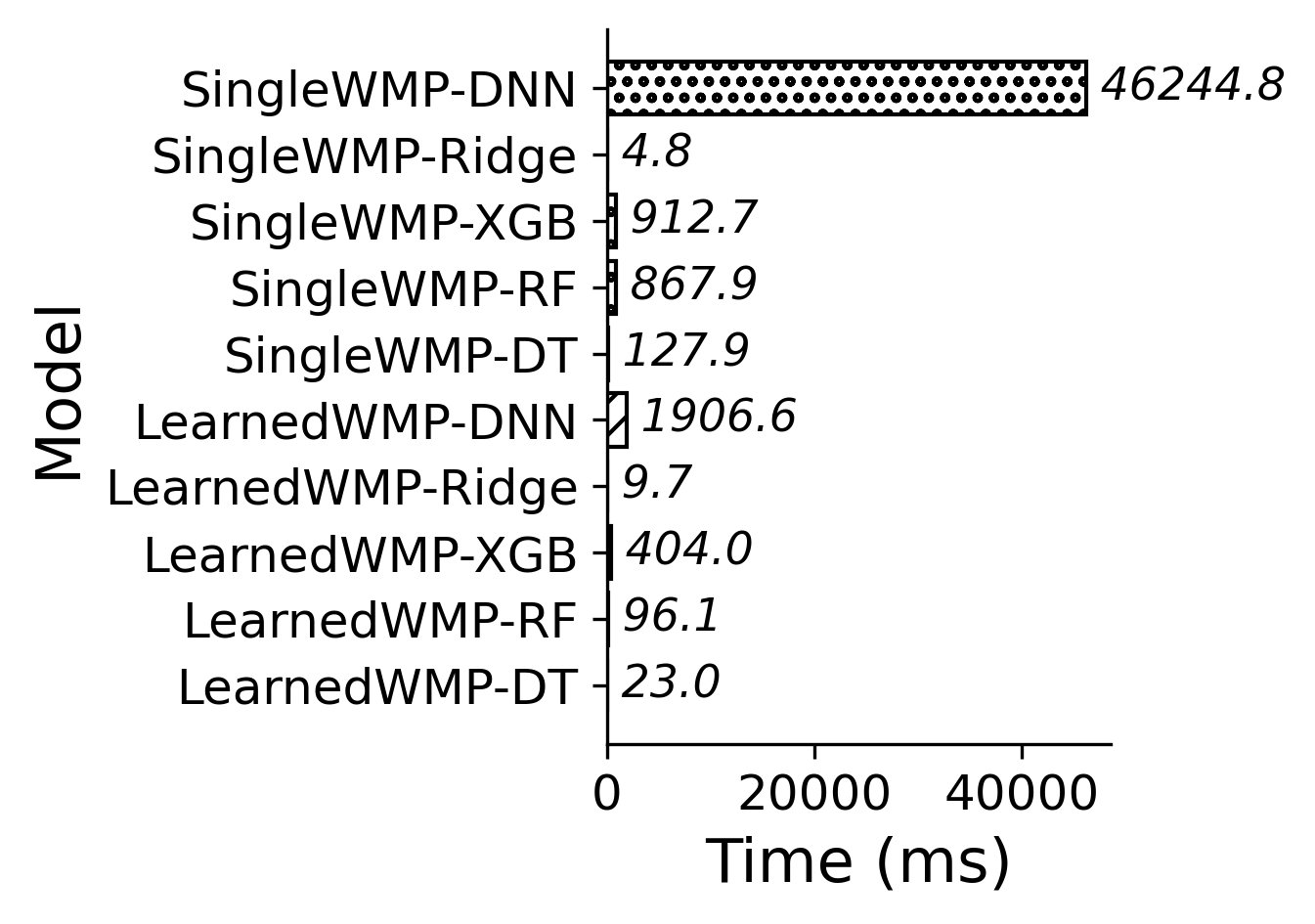}
         \caption{\texttt{TPC-DS}}
         \label{fig:TPCDS-mtl}
     \end{subfigure}
     \hfill
     \begin{subfigure}[b]{0.31\textwidth}
         \centering
         \includegraphics[width=\textwidth]{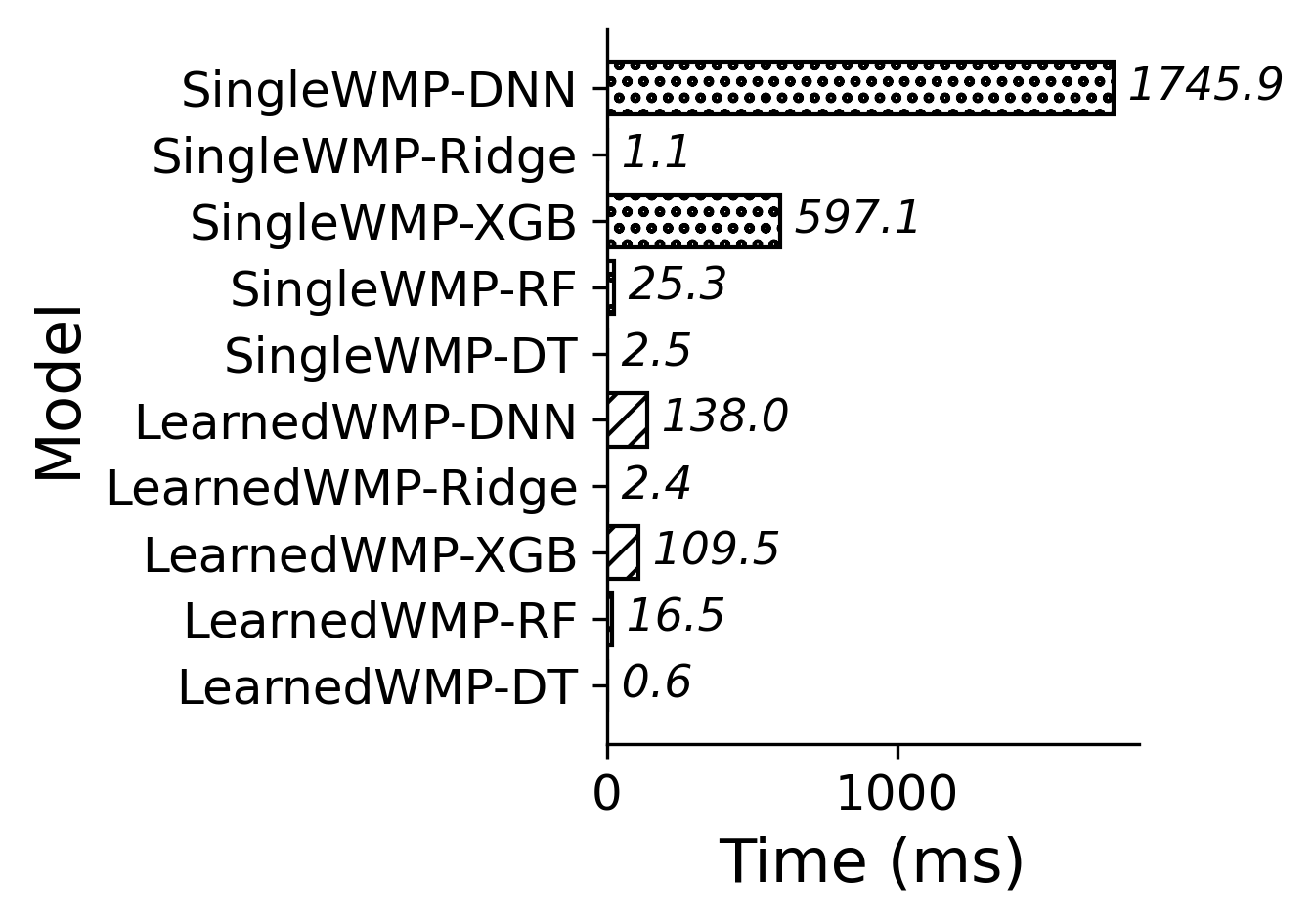}
         \caption{\texttt{JOB}}
         \label{fig:JOB-mtl}
     \end{subfigure}
    \hfill
     \begin{subfigure}[b]{0.31\textwidth}
         \centering
         \includegraphics[width=\textwidth]{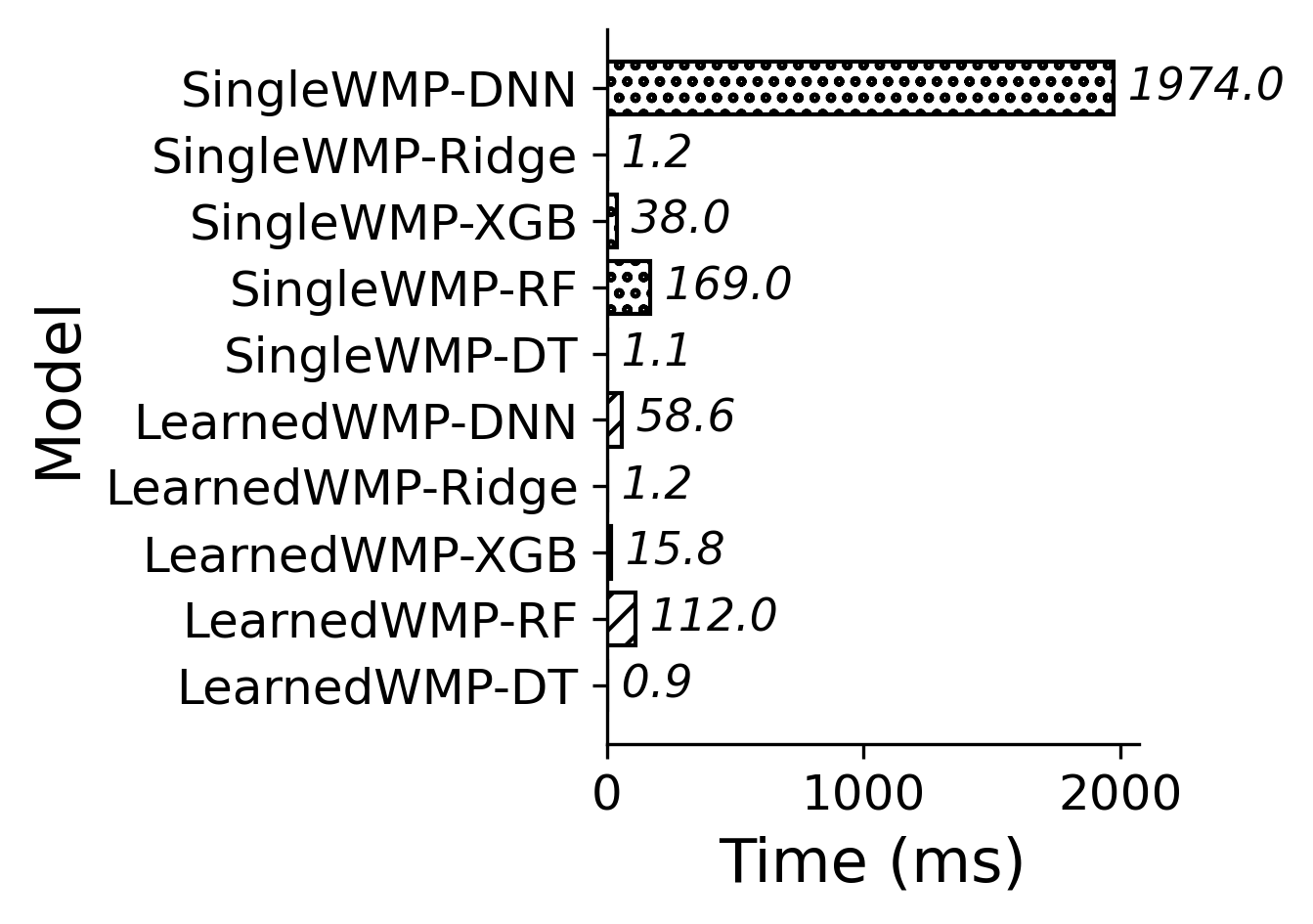}
         \caption{\texttt{TPC-C}}
         \label{fig:TPC-C-mtl}
     \end{subfigure}
     }
        \caption{ML model training time}
        \label{fig:model-training-latency}
        \vspace{-10px}
\end{figure*}

\begin{figure*}[t!]
\makebox[\linewidth][c]{%
     \centering
     \begin{subfigure}[b]{0.31\textwidth}
         \centering
         \includegraphics[width=\textwidth]{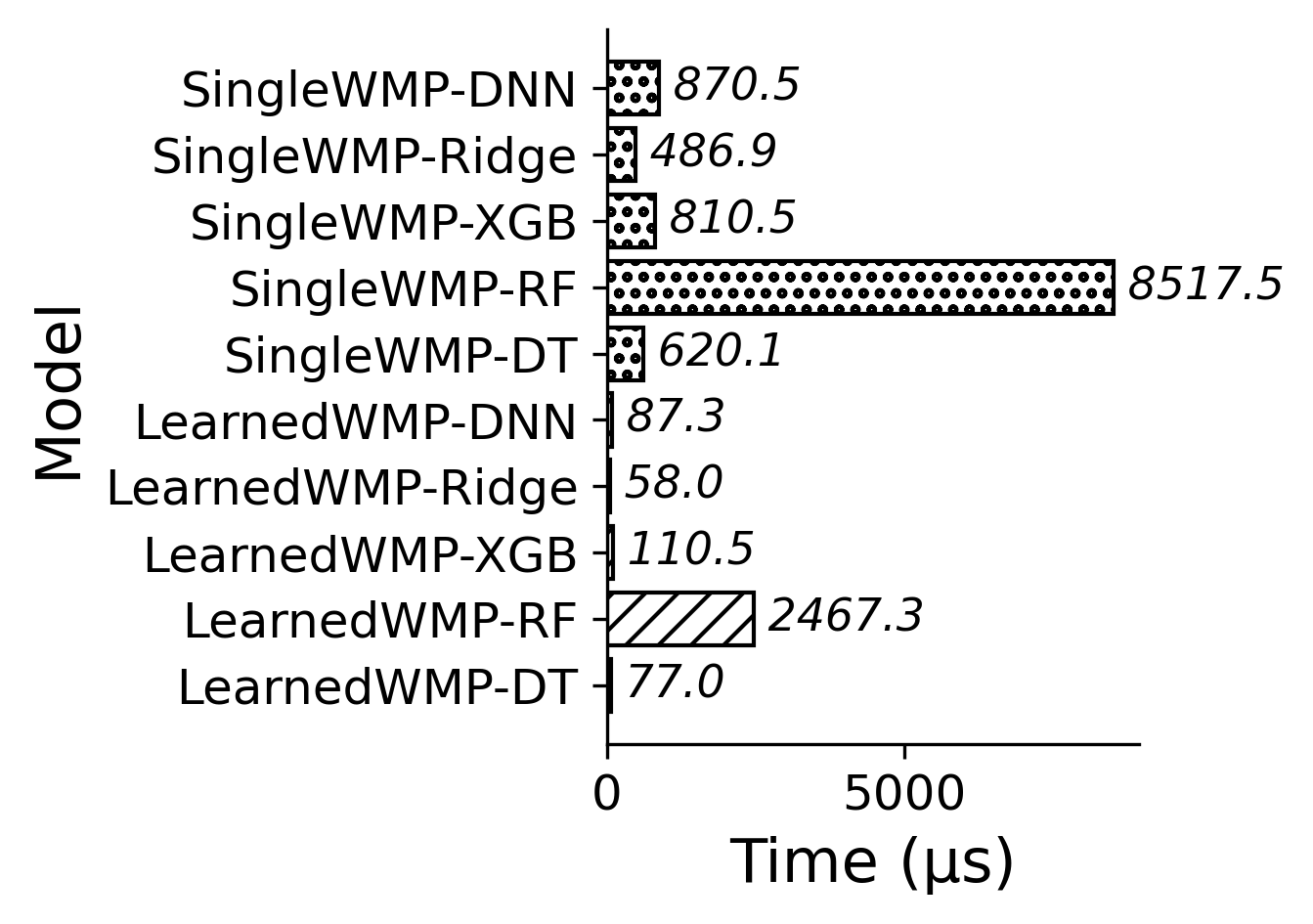}
         \caption{\texttt{TPC-DS}}
         \label{fig:TPC-DS-mil}
     \end{subfigure}
     \hfill
     \begin{subfigure}[b]{0.31\textwidth}
         \centering
         \includegraphics[width=\textwidth]{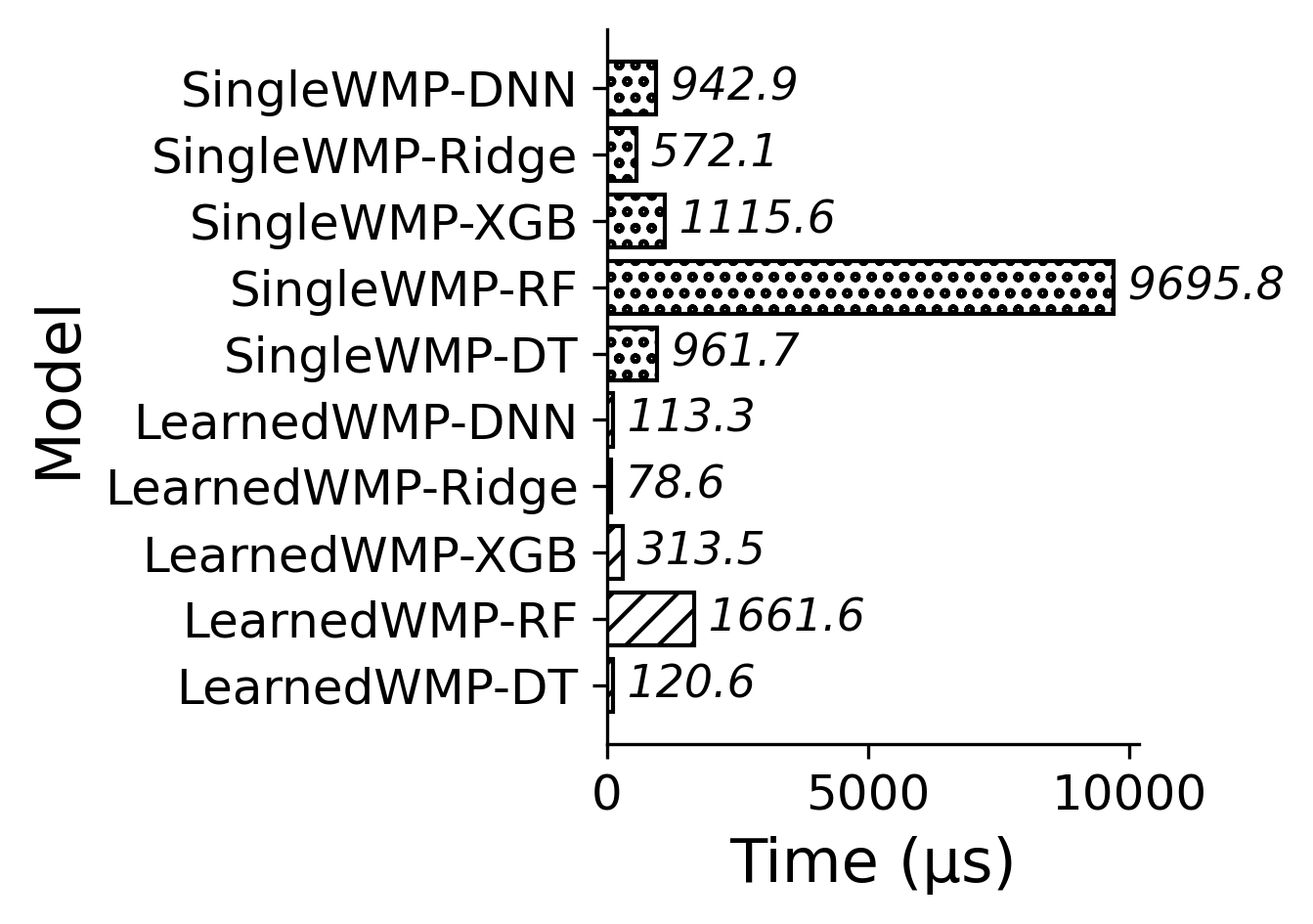}
         \caption{\texttt{JOB}}
         \label{fig:JOB-mil}
     \end{subfigure}
     \hfill
     \begin{subfigure}[b]{0.31\textwidth}
         \centering
         \includegraphics[width=\textwidth]{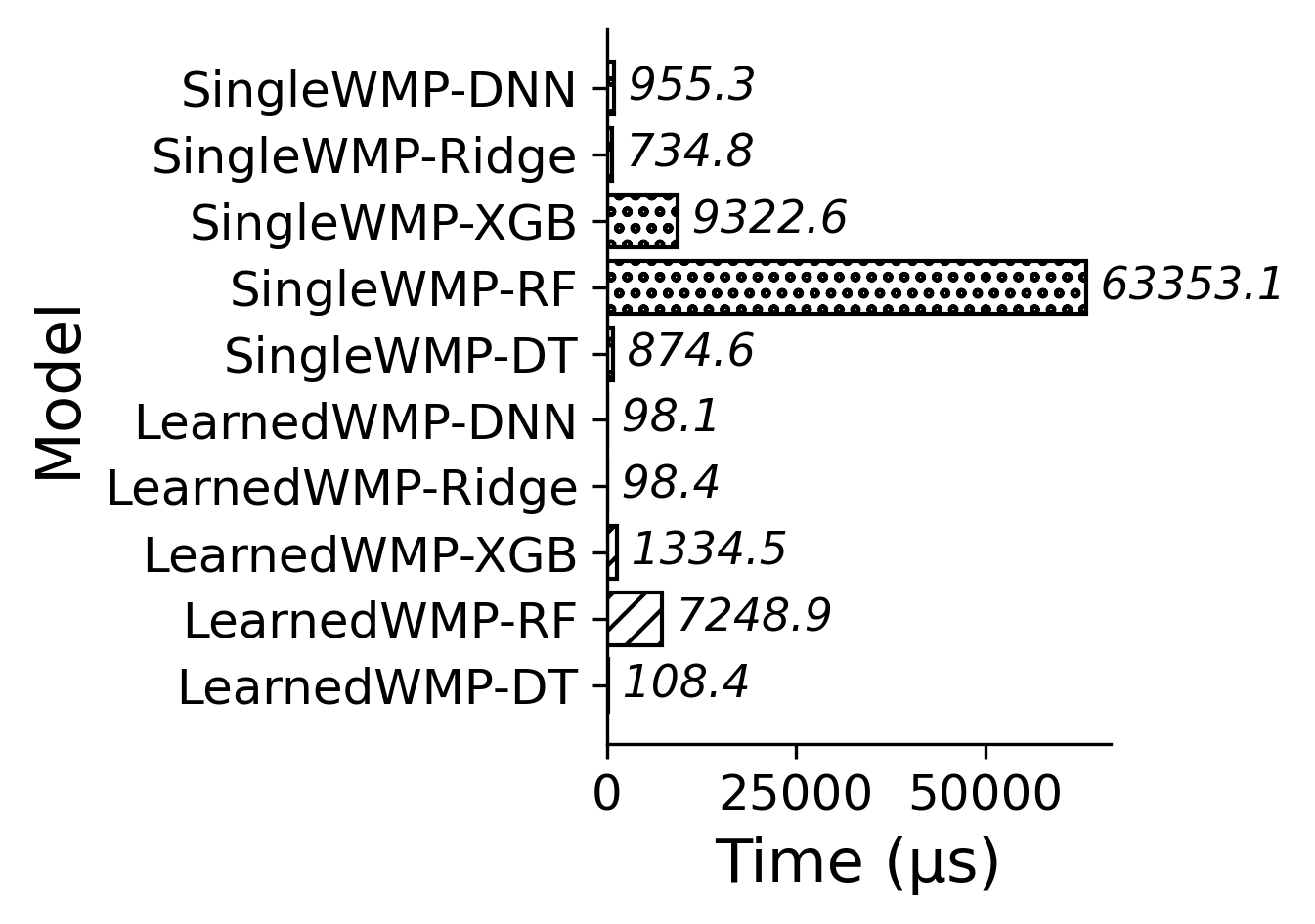}
         \caption{\texttt{TPC-C}}
         \label{fig:TPC-C-mil}
     \end{subfigure}
     }
        \caption{ML model inference time}
        \label{fig:model-inference-latency}
    \vspace{-10px}
\end{figure*}

\begin{figure*}
\makebox[\linewidth][c]{%
     \centering
     \begin{subfigure}[b]{0.31\textwidth}
         \centering         \includegraphics[width=\textwidth]{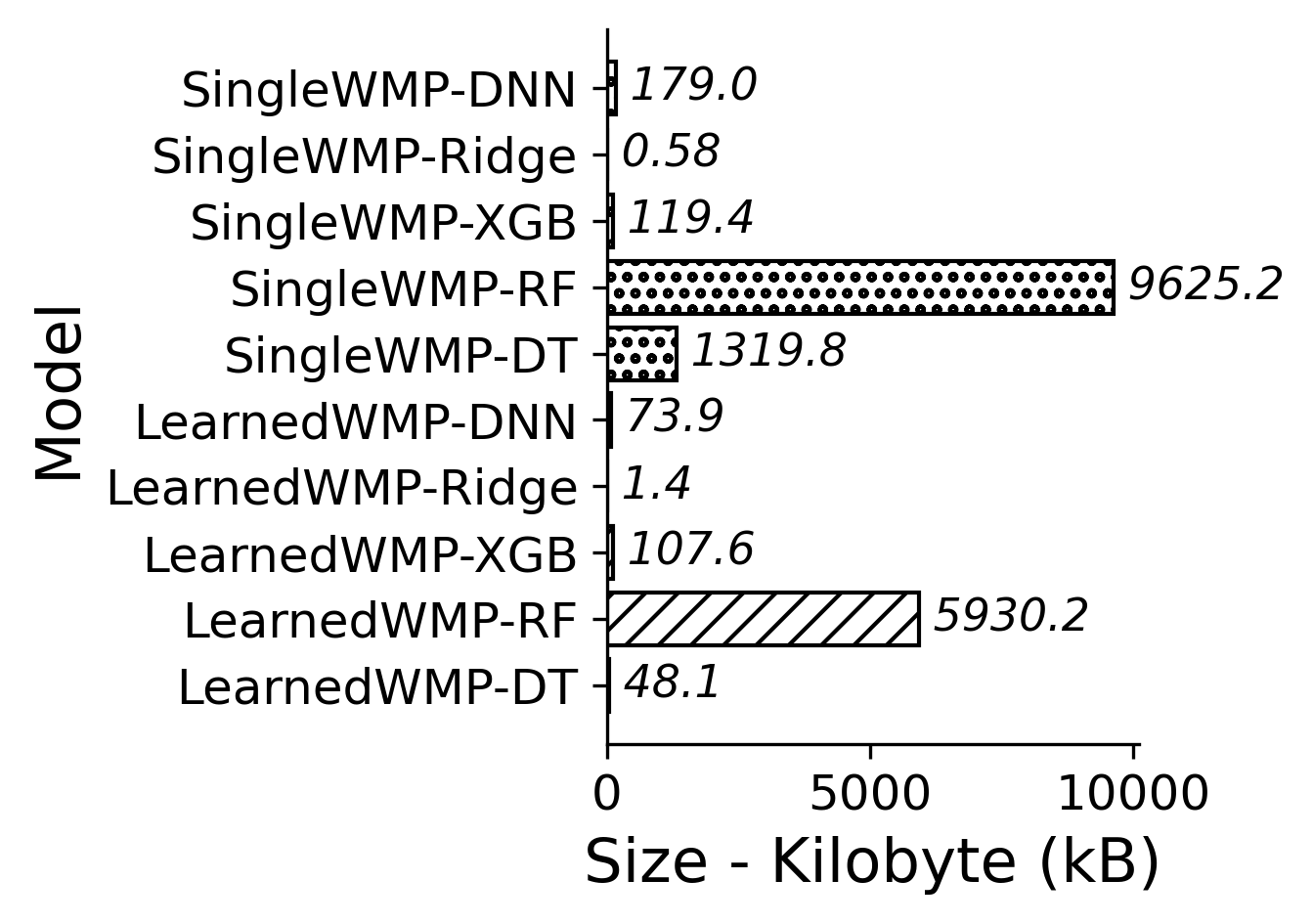}
         \caption{\texttt{TPC-DS}}
         \label{fig:TPC-DS-ms}
     \end{subfigure}
     \hfill
     \begin{subfigure}[b]{0.31\textwidth}
         \centering
         \includegraphics[width=\textwidth]{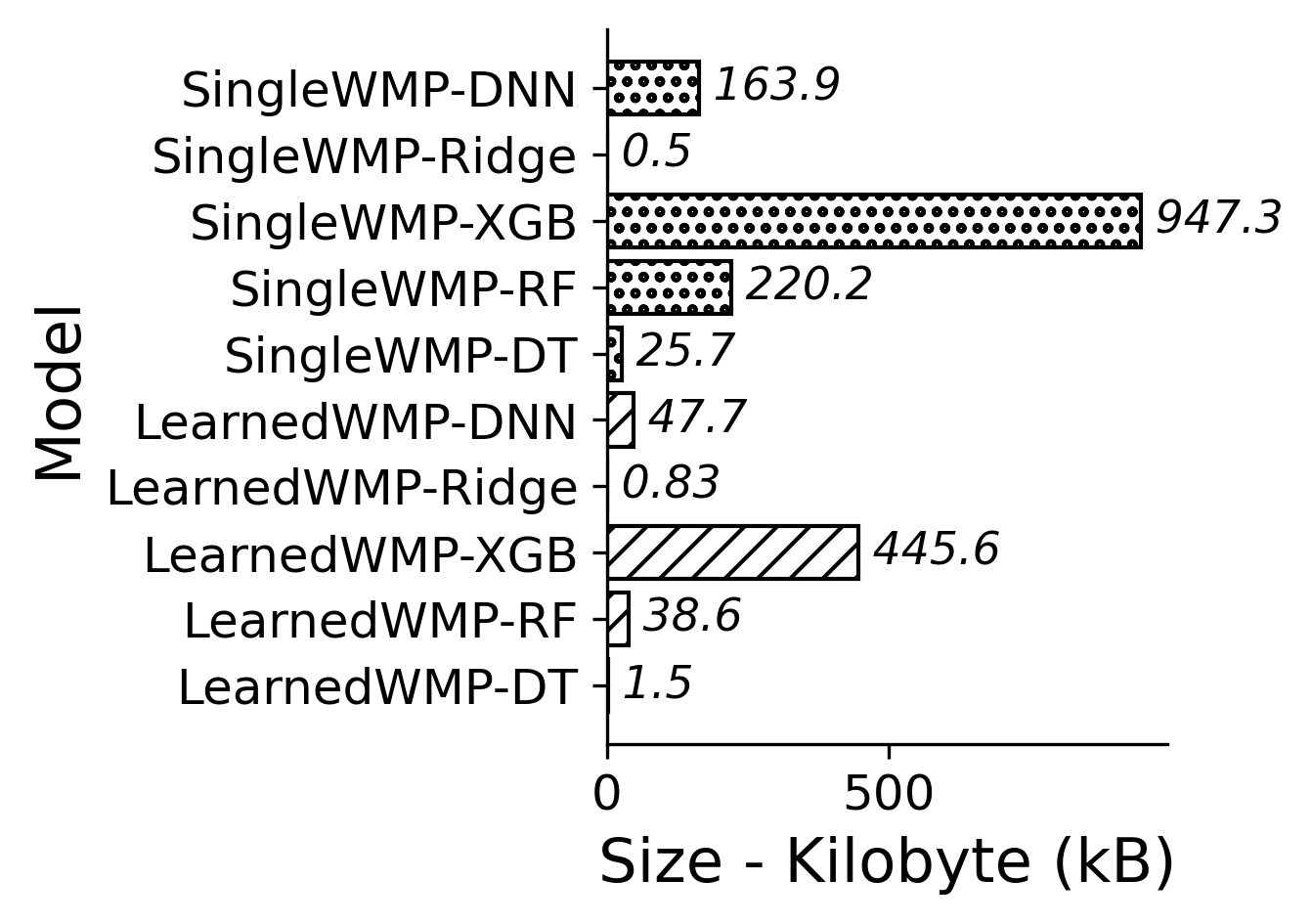}
         \caption{\texttt{JOB}}
         \label{fig:JOB-ms}
     \end{subfigure}
     \hfill
     \begin{subfigure}[b]{0.31\textwidth}
         \centering
         \includegraphics[width=\textwidth]{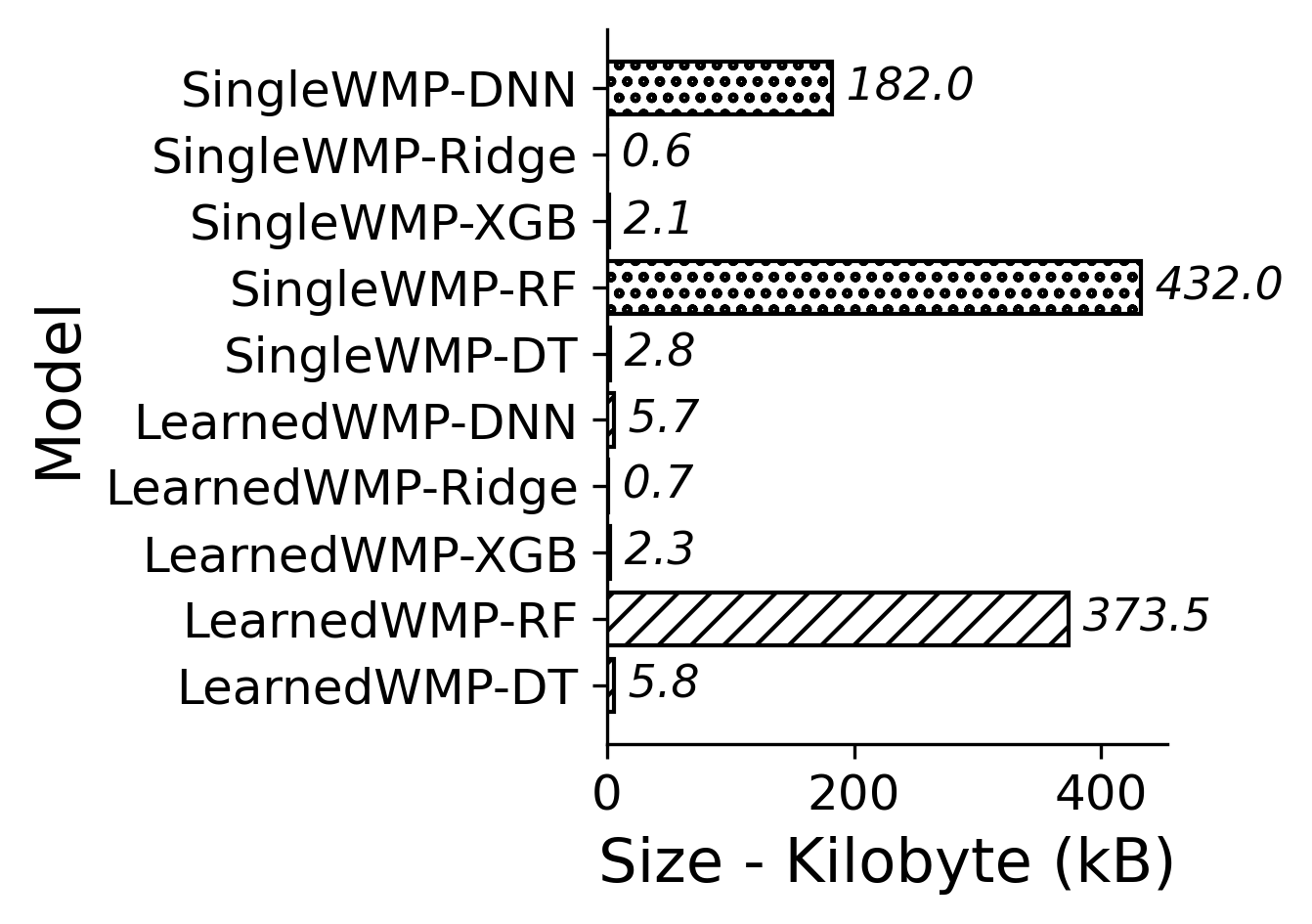}
         \caption{\texttt{TPC-C}}
         \label{fig:TPC-C-ms}
     \end{subfigure}
     }
        \caption{ML model size}
        \label{fig:model-size}
    \vspace{-10px}
\end{figure*}

Fig.~\ref{fig:model-inference-latency} compares the inference time of SingleWMP-based and LearnedWMP-based models. The LearnedWMP-based models achieved between 3x and 10x acceleration compared to their equivalent SingleWMP-based models. As an example, for inference of TPC-DS workloads, LearnedWMP-DNN took 87.3 µs as compared to 870.5 µs needed by SingleWMP-DNN. Similarly, with JOB, LearnedWMP-XGB needed 313.3 µs for inference, while SingleWMP-XGB took 1115.6 µs.
Accelerated training and inference performance of the LearnedWMP models can be attributed to our approach of formulating the training and inference task at the level of workloads, not at the level of individual queries. LearnedWMP-based models process batches of queries at the same time and, therefore, speed up the computation during both training and inference. In contrast, SingleWMP-based models require a longer time for training and inference as they work one query at a time.

\stitle{Model Size.} The model size largely depends on the learning algorithm and the feature space complexity of the training set. Fig. \ref{fig:model-size} shows the size of LearnedWMP-based and SingleWMP-based models. LearnedWMP-based models were significantly smaller when compared with equivalent SingleWMP-based models. For example, when compared with SingleWMP-DNN, LearnedWMP-DNN is 59\% TPC-DS, 72\% JOB, and 97\% TPC-C smaller than SingleWMP-DNN. We see a similar pattern with XGBoost, RF, and DT when comparing their LearnedWMP-based models with equivalent SingleWMP-based models.
Batching training queries as workloads compress the information a LearnedWMP model needs to process during training. This compression helps the LearnedWMP approach produce smaller models as compared with models of the SingleWMP approach. Ridge is an exception to this observation. The size of a LearnedWMP-Ridge model is larger than its equivalent SingleWMP-Ridge model. This was expected as Ridge learns a set of coefficients, one for each input feature in the training dataset. In our training datasets, each LearnedWMP training example has more input features, one per query template, than the number of features in a SingleWMP training example. As a result, LearnedWMP-Ridge learns more coefficients during training and produces larger models.

\subsection{Sensitivity Analysis}
\label{sec:sensitivity}

In the next set of experiments, we investigate the impact
of various parameters of LearnedWMP, such as the batch size parameter $s$, the number of query templates $K$, the choice of the learning query templates techniques, and their effect on the memory prediction accuracy

\stitle{Learning Query Templates.} In the first phase of LearnedWMP, queries are assigned to templates based on their similarity in query plan characteristics and estimates, with the expectation that queries in the same template exhibit similar memory usage (See Section\ref{subsection:learning-query-templates}). We evaluated our method against four other approaches for learning query templates:
\begin{enumerate}
    \item \textbf{Query plan based}: Our proposed LearnedWMP model assigns queries to query templates by extracting features from the query plan and then employing a $k$-means clustering algorithm. Details can be found in section \ref{subsection:learning-query-templates}.
    \item \textbf{Rule based}: We create a set of rules, one per template, to classify a query statement into one of the pre-defined query templates. Subject matter experts, such as DBAs, may need to be involved in defining these rules \cite{zhou2020query}.
    \item \textbf{Bag of Words based}: We extract unique keywords from the entire training query corpus to build a vocabulary. Each query expression generates a feature vector representing the count of each vocabulary word in the query. The $k$-means clustering algorithm then assigns these feature vectors to different query templates.
    \item \textbf{Text mining based}: This is a variation of the bag of words approach, which indiscriminately extracts unique keywords from the training corpus. In contrast, in this approach, the vocabulary includes only those keywords that are either database object names (e.g., a Table name) or SQL clauses (e.g., group by). Other keywords are ignored. After vocabulary building, we apply similar steps to the bag of words approach to generate a feature vector for each query and then apply $k$-means clustering to assign them to templates.
    \item \textbf{Word embeddings based}: 
    Word Embeddings address two limitations of bag-of-words methods: dealing with numerous keywords and capturing keyword proximity \cite{chollet2021deep}. Using word embeddings, With word embeddings, we construct a vocabulary from the training query corpus and generate a feature vector for each query expression. Applying $k$-means clustering assigns these feature vectors to templates.

\end{enumerate}

\noindent To evaluate the performance of the five alternative methods for learning templates, we used the LearnedWMP-XGB model with JOB workloads. We trained five LearnedWMP-XGB models, each using a different method for learning templates. Fig. \ref{fig:templates} compares the accuracy of these five models. The model--- labeled query plan (ours) in the figure---that uses our original method for learning query templates outperformed the four alternatives. 
Compared with the alternatives, the LearnedWMP method uses features from the query plan. The query plans include estimates that are strong indicators of the resource usage of the queries. A prior research \cite{ganapathi2009predicting} made a similar observation. In contrast, the alternative methods extract features directly from the query expression, which does not provide insights into the query's memory usage. A major limitation of the rules-based method is that coming up with effective rules requires the knowledge of human experts, which can be both a difficult and a slow process.

\begin{figure}[t]
  \includegraphics[width=0.45\textwidth]{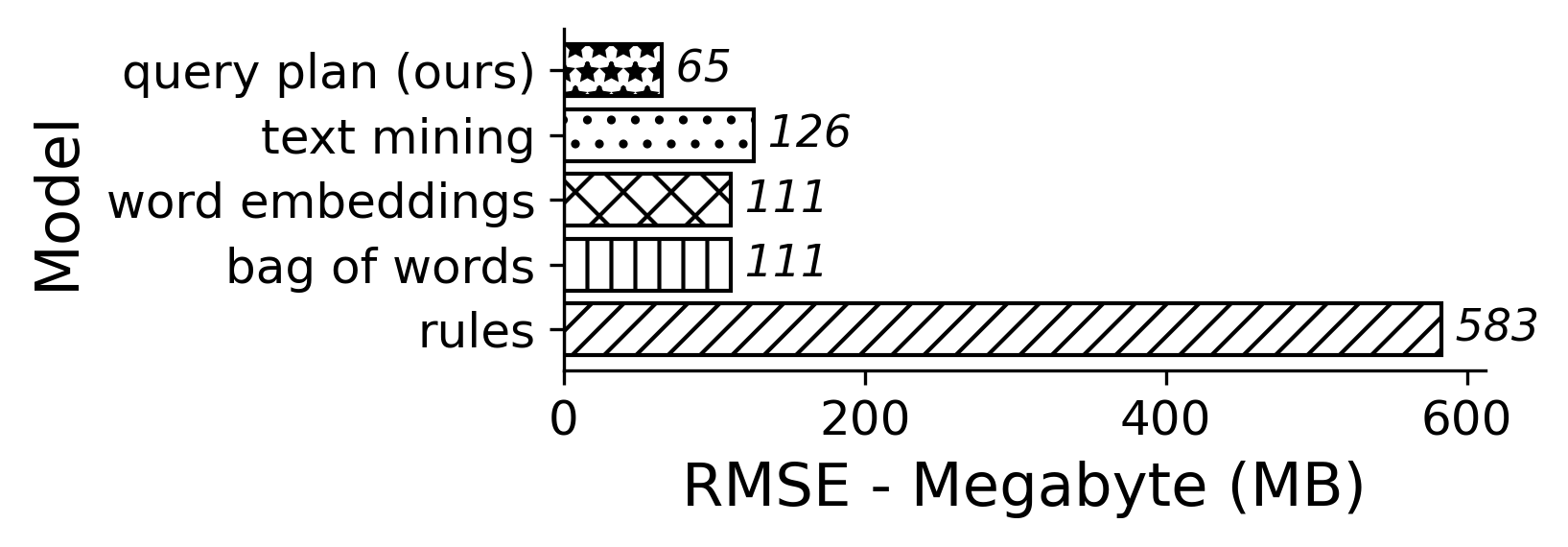}
  \caption{LearnedWMP's accuracy performance achieved by different methods for learning query templates.}
  \label{fig:templates}
  \vspace{-10px}
\end{figure}

\begin{figure*}[t]
\makebox[\linewidth][c]{%

\begin{subfigure}[b]{.33\textwidth}
\centering
\includegraphics[width=.90\textwidth]{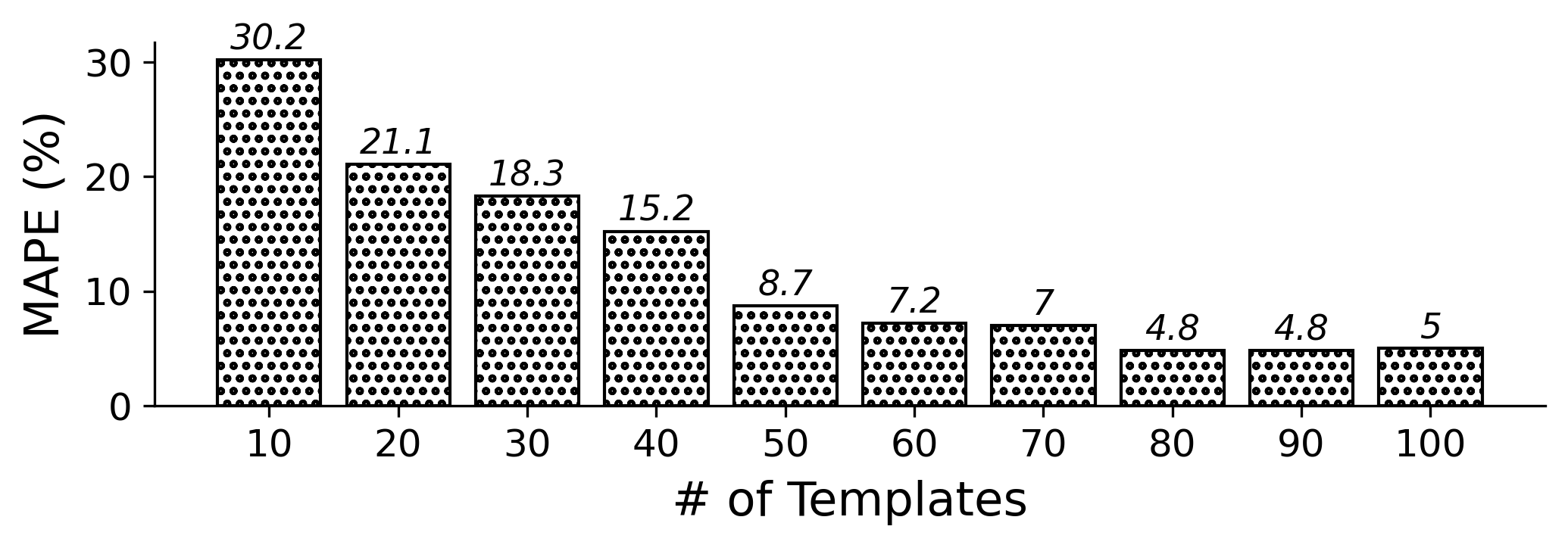}
\caption{\textbf{TPC-DS}}
\end{subfigure}
\hfill
\begin{subfigure}[b]{0.33\linewidth}
\centering
\includegraphics[width=.90\textwidth]{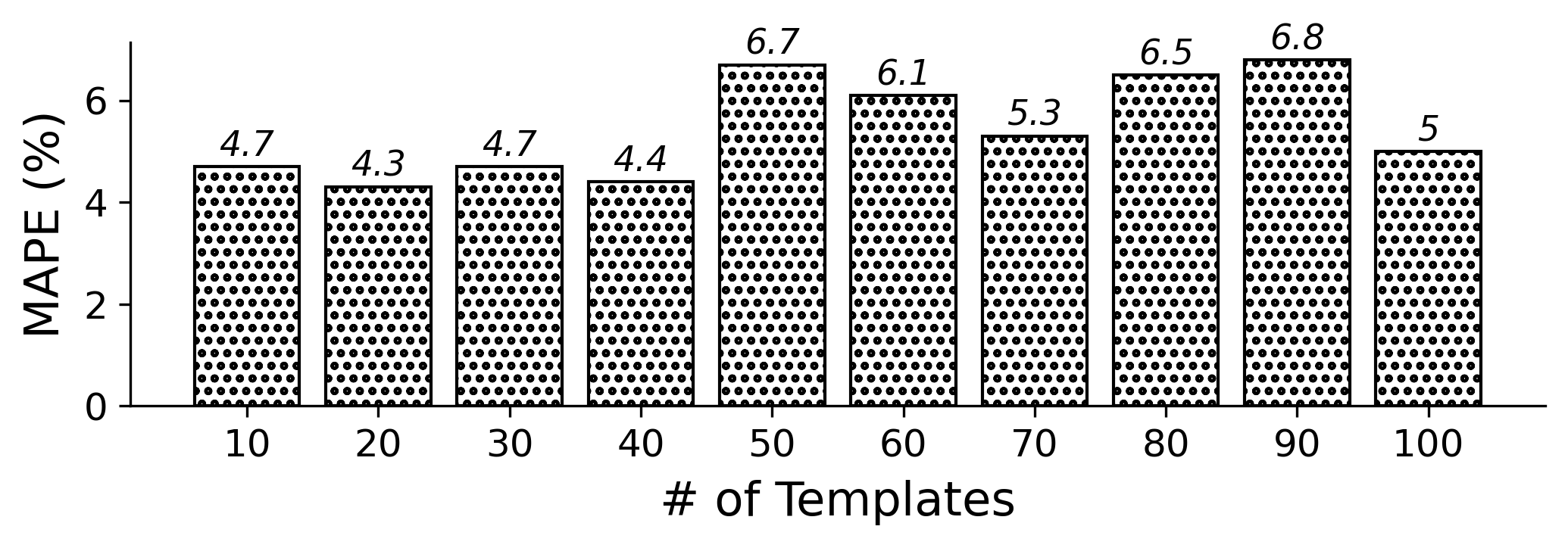}
\caption{\textbf{JOB}}
\end{subfigure}%
\hfill
\begin{subfigure}[b]{.33\textwidth}
\centering
\includegraphics[width=.90\textwidth]{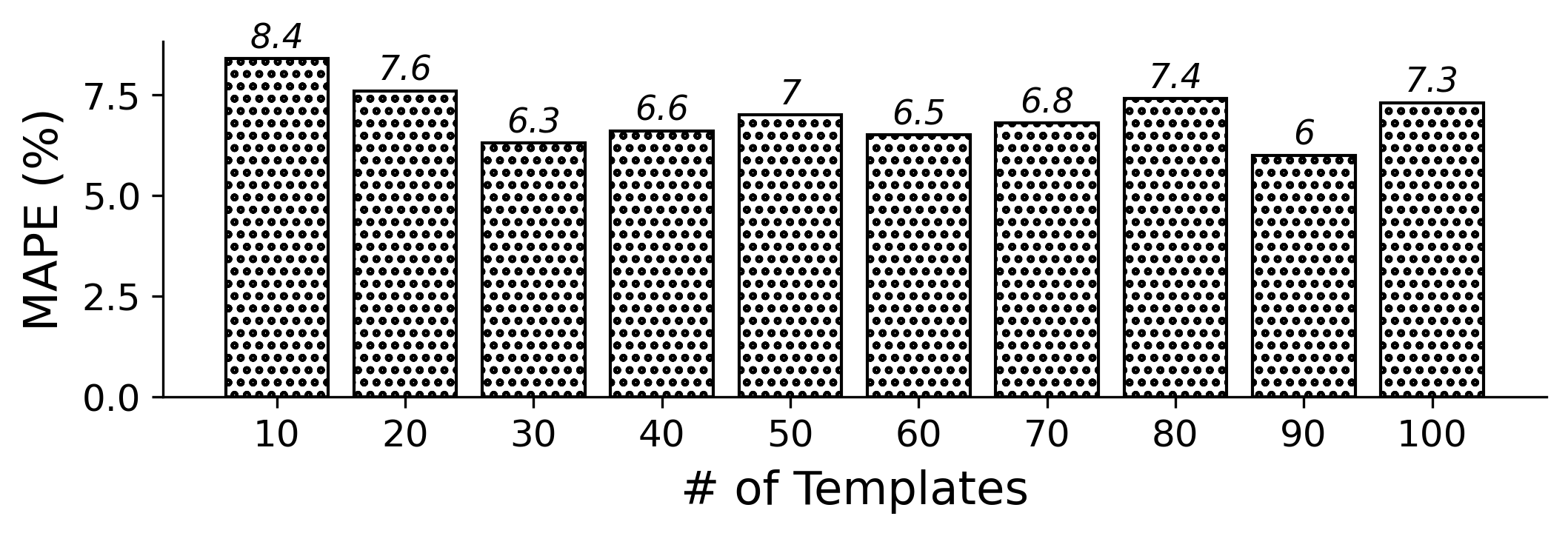}
\caption{\textbf{TPC-C}}
\end{subfigure}
}
\caption{MAPE at different template sizes with LearnedWMP XGBoost Model for the three evaluation datasets (a) TPC-DS, (b) JOB, and (c) TPC-C}
\label{fig:templatesizes}
\vspace*{-10px}
\end{figure*}


\stitle{Effect of the number of query templates.} As described in section \ref{subsection:learning-query-templates}, LearnedWMP assigns queries into templates, such as queries with similar query plan characteristics in the same groups. This experiment aimed to assess the effects of the number of templates on the LearnedWMP model's accuracy. In the experiment, we tested the performance of the LearnedWMP-XGB model on three datasets using 10 to 100 templates, comparing the model's performance across different template sizes. 

\noindent We used the Mean Absolute Percent Error (MAPE)~\cite{de2016mean} to evaluate and compare these models, each using a different number of templates.  
 The scale of the error can vary significantly when changing the number of templates. MAPE  is unaffected by changes in the error scale when comparing models trained with different numbers of templates because it calculates a relative error. 
We used equation (\ref{eqmse}) to compute the MAPE of the LearnedWMP-XGB model. 
\begin{align}
    MAPE = \frac{1}{N}\sum_{i=1}^N{\frac{|y_i-\hat{y_i}|}{y_i}} \times 100
    \label{eqmse}
\end{align}
We computed the relative estimation error between the actual and predicted memory usage by dividing the absolute difference between them by the actual value and then averaged the relative estimation errors across all test workloads. The resulting average was multiplied by 100 to obtain the MAPE, ranging from 0 to 100 percent. Fig. \ref{fig:batchsizes} shows the performance of LearnedWMP-XGB as a factor of the number of templates. Fig~\ref{fig:templatesizes} shows the performance of the LearnedWMP-XGB model for each template size for the three datasets. In the case of the TPC-DS dataset, performance improved gradually with an increasing number of templates. The best performance was observed at 100 templates. However, for the JOB and TPC-C datasets, performance varied as the number of templates increased, with optimal performance achieved within the range of 20 to 40 templates. 
We argue that this correlation between the number of queries and the optimal number of templates is due to the characteristics of each dataset. The larger TPC-DS dataset benefits from a greater diversity of queries (93,000) queries generated from (99) templates, which allows clustering with a higher number of templates. However, the smaller JOB and TPC-C datasets do not possess the same level of query variation, which explains why the best performance was achieved with a moderate number of templates. 

\stitle{Effect of the batch size parameter} The experiments we discussed so far used a constant workload batch size of 10. The batch size, $s$, is a tunable hyperparameter of the LearnedWMP model. We tried different workload batch sizes and compared their impact on the LearnedWMP model's accuracy. We used the TPC-DS dataset and the LearnedWMP-XGB model for this experiment. We used 12 values for the batch size: [1, 2, 3, 5, 10, 15, 20, 25, 30, 35, 40, 45, 50]. For each value, we created TPC-DS training and test workloads, which we used for training and evaluating a LearnedWMP-XGB model. We computed and used MAPE to compare the relative accuracy performance of these models. Fig. \ref{fig:batchsizes} shows the performance of LearnedWMP-XGB as a factor of the batch size. We can see that with increasing batch size, the accuracy of the memory estimation improves. 
The improvement is more rapid initially, gradually slowing down, which is expected of any learning algorithm as it reaches the perfect prediction. For example, at batch size 2, the estimation error was 10.4\%. At batch size 10, the error was reduced to 3.8\%. We have seen a similar improvement in prediction accuracy with the remaining two experimental datasets. This observation substantiates our position that batch estimation of workload memory is more accurate than estimating one query's memory at a time. Additionally, we compared the MAPE of the LearnedWMP model with batch size 1 with the SingleWMP model's MAPE. At batch size 1, LearnedWMP's MAPE is 10.2, whereas SingleWMP's MAPE is 3.6. 
The SingleWMP model outperformed the LearnedWMP batch 1 model, which was expected since SingleWMP was directly trained with query plan features, providing strong signals for individual query memory usage. In contrast, the LearnedWMP model mapped queries into templates and generated predictions based on collective memory usage, lacking the ability to learn from individual query features. While LearnedWMP had weaker signals for single query predictions, it demonstrated higher accuracy than SingleWMP when generating predictions for query batches. However, as we have seen in \ref{sec:learnedwmp-accuracy}, LearnedWMP generates more accurate predictions than SingleWMP when generating predictions for batches of queries.

\section{Related Work}
\label{sec:related}

\noindent Our research relates to ML methods for (i) database query optimization \cite{jarke1984query, ioannidis1996query, chaudhuri1998overview}, (ii) database query resource estimation \cite{li2012robust}, (iii) query-based workload analysis \cite{ma2018query}, and (iv) distribution regression problems. Each of these areas has extensive research literature, and we discuss some of the most significant ones. 

\stitle{ML for Database Query Optimization} 
In the broader query optimization topic, besides query resource estimation, many recent works explored ML techniques to learn different tasks related to query optimization. Some of the key tasks include cardinality estimation \cite{han2021cardinality, kim2022learned, liu2015cardinality}, query latency prediction \cite{zhou2020query, akdere2012learning}, index selection \cite{sharma2018case, ding2019ai}. A recent cardinality estimation benchmark \cite{han2021cardinality} evaluated eight ML-based cardinality estimation methods---MSCN, LW-XGB, LW-NN, UAE-Q, NeuroCard, BayesCard, DeepDB, and FLAT. Zhou et al. \cite{zhou2020query} proposed a graph-based deep learning method to predict the execution time of concurrent queries. Akdere et al. \cite{akdere2012learning} used support vector machine (SVM) and linear regression to predict the execution time of analytical queries. Marcus and Papaemmanouil \cite{marcus2019plan} proposed a plan-structured neural network architecture, which uses custom neural units designed at the level of query plan operators to predict query execution time. Ahmad et al. \cite{ahmad2011predicting} proposed an ML-based method for predicting the execution time of batch query workloads. They relied upon DBAs to define a set of query types used to create simulated workloads and model interactions among queries. Contender \cite{duggan2014contender} is a framework for predicting the execution time of concurrent analytical queries that compete for I/O.  Ding et al. \cite{ding2019ai} applied classification techniques to compare the relative cost of a pair of query plans and use that insight in index recommendations.

\begin{figure}[t]
  \includegraphics[width=0.45\textwidth]{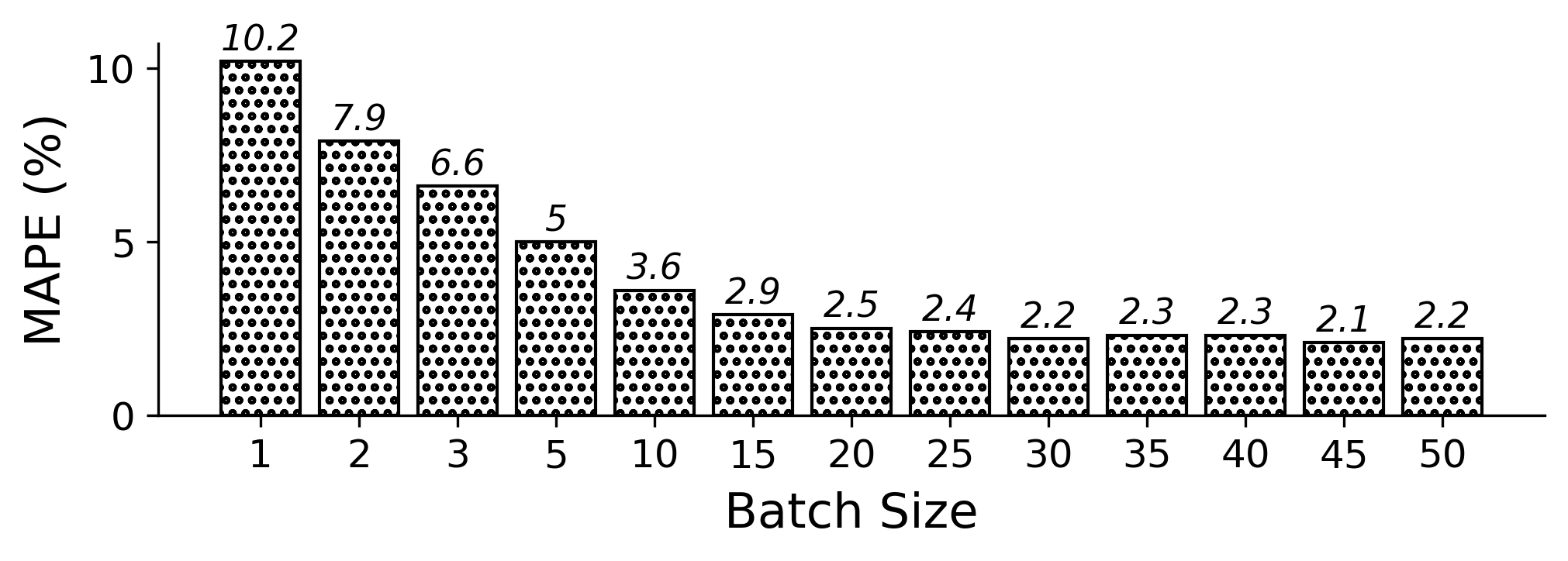}
  \caption{TPC-DS MAPE at different batch sizes with LearnedWMP XGBoost Model.}
  \label{fig:batchsizes}
  \vspace{-10px}
\end{figure}


\balance
\stitle{ML for Database Query Resource Estimation} Closer to our problem are methods that estimate computing resources---such as memory, CPU, and I/O---for executing database queries. Tang et al. used ML to classify queries into low, medium, and high resource consumption. They employed separate models for memory and CPU, utilizing a bag-of-words approach to extract query keywords as input features. XGBoost proved the most accurate among the three compared algorithms\cite{tang2021forecasting}.
Ganapathi et al. \cite{ganapathi2009predicting} tried five ML techniques to predict run-time resource consumption metrics of individual queries. They achieved the best results with the kernel canonical correlation analysis (KCCA) algorithm. Li et al. \cite{li2012robust} applied boosted regression trees to predict individual queries' CPU and I/O costs. They modeled the resource requirements of each database operator separately by computing a different set of features for each operator type. All the above three approaches \cite{tang2021forecasting, ganapathi2009predicting, li2012robust} generate resource predictions at the level of individual queries. In contrast, our proposed method estimates memory at the level of workload query batches, which we found more accurate and computationally efficient. 

\stitle{ML for Query-based Workload Analysis} There is a line of research that considers query-based workload analysis. Higginson et al. \cite{higginson2020database} have applied time series analysis, using ARIMA and Seasonal ARIMA (SARIMA), on database workload monitoring data to identify patterns such as seasonality (reoccurring patterns), trends, and shocks. They used these time series patterns in database capacity planning. Kipf et al.\cite{kipf2018learned} use Multi-Set Convolutional Network (MSCN) for workload cardinality estimation, representing query features as sets of tables, joins, and predicates. Similar methods have been used for constructing query-based cardinality estimators (e.g., .\cite{negi2023robust}]. 
However, these approaches involve significant sampling overhead or rely on static data featurization, unsuitable for modern databases. Additionally, they lack explanations for the relationships between data, queries, and actual cardinalities. DBSeer \cite{mozafari2013performance} employs ML to predict resource metrics in OLTP workloads, using DBSCAN for transaction type learning and linear regression, decision trees, and neural networks for CPU, I/O, and memory demand prediction. In contrast, our approach does not cluster transactions or use query expressions for clustering. Instead, we learn query templates based on simple features in the query plan, showing a stronger correlation with runtime memory usage. In our experiments, we compared DBSCAN-based templates with $k$-means and found the latter more accurate for resource prediction.
 
\stitle{ML for Distribution Regression Problems} Distribution regression has emerged as a popular ML approach for mapping complex input probability distributions to real-valued responses, particularly in supervised tasks that require handling input and model uncertainty~\cite{law2018bayesian}, emerging as a promising alternative to traditional techniques like random forests and neural networks \cite{li2019deep}. To the best of our knowledge, we are the first to apply distribution regression to model resource demand forecasting of database workloads. Outside the database domain, some of the illustrative use cases of distribution regression include predicting health indicators from a patient's list of blood tests \cite{mao2022coefficient}, solar energy forecasting, and traffic prediction\cite{li2019deep}. Many recent papers have offered approaches and optimization techniques for solving distribution regression tasks (e.g., \cite{law2018bayesian,li2019deep,mao2022coefficient }).

\section{Conclusions}
\label{sec:conclusions}

\noindent 

We proposed a novel approach to predicting memory usage of database queries in batches.  Our approach is a paradigm shift from the state of the practice and the state-of-the-art methods designed to estimate resource demand for single queries. As an embodiment of our approach, we presented LearnedWMP, a method for estimating the working memory demand of a batch of queries, a workload.
The LearnedWMP method operates in three phases. First, it learns query templates from historical queries. Second, it constructs histograms from the training workloads. Third, using training workloads, it trains a regression model to predict the memory requirements of unseen workloads. We model the prediction task as a distribution regression problem. We performed a comprehensive experimental evaluation of the LearnedWMP model against the state-of-the-practice method of a contemporary DBMS, multiple sensible baselines, and state-of-the-art methods. Our analysis demonstrates that our proposed method significantly improves the memory estimation of the current state of the practice. Additionally, LearnedWMP matches the performance of advanced ML-based methods trained with a single-query approach. It generates smaller models, enabling faster training and quicker memory usage predictions. We conducted parameter sensitivity analysis and explored various strategies for learning query templates from historical DBMS queries. Our novel LearnedWMP model presents an alternative perspective on a crucial DBMS problem, easily integratable with major DBMS products.
\balance
\newpage


\bibliographystyle{abbrv}
\bibliography{bibliography}

\end{document}